\documentclass[12pt,oneside,english]{amsart}
\usepackage{color}
\usepackage{babel}
\usepackage{units}
\usepackage{amstext}
\usepackage{amsthm}
\usepackage{amssymb}
\usepackage{paralist}
\usepackage{setspace,graphicx,tikz}
\usetikzlibrary{positioning}
\usepackage[T1]{fontenc}

\usepackage[authoryear]{natbib}
\usepackage[unicode=true,pdfusetitle,
 bookmarks=true,bookmarksnumbered=false,bookmarksopen=false,
 breaklinks=false,pdfborder={0 0 0},pdfborderstyle={},backref=false,colorlinks=true]
 {hyperref}
\hypersetup{
 pdfborderstyle={},pdfborderstyle={},urlcolor=blue,linkcolor=blue,citecolor=blue}

\makeatletter
\numberwithin{figure}{section}
\theoremstyle{plain}
\newtheorem{fact}{\protect\factname}
\newtheorem{thm}{\protect\theoremname}
\newtheorem{cor}{\protect\corollaryname}

\newtheorem{lem}{\protect\lemmaname}

\usepackage{babel}
\usepackage{babel}
\usepackage{fullpage}
\usepackage[foot]{amsaddr}
\setcitestyle{round}

\makeatletter
\renewenvironment{proof}[1][\proofname] {\par\pushQED{\qed}\normalfont\topsep6\p@\@plus6\p@\relax\trivlist\item[\hskip\labelsep\bfseries#1\@addpunct{.}]\ignorespaces}{\popQED\endtrivlist\@endpefalse}
\makeatother

\makeatother

\providecommand{\corollaryname}{Corollary}
\providecommand{\factname}{Fact}
\providecommand{\lemmaname}{Lemma}
\providecommand{\propositionname}{Proposition}
\providecommand{\theoremname}{Theorem}
\providecommand{\claimname}{Claim}
\usepackage{soul}
\begin{document}
\global\long\def\eu{u_{\theta}}%
 
\global\long\def\act{\alpha}%
 
\global\long\def\actmin{\alpha^{\min}}%
 
\global\long\def\ra{r_{aut}}%
 
\global\long\def\ru{r_{uni}}%
 
\global\long\def\rmax{r_{bnd}}%
\global\long\def\rg{r_{grp}}%

\definecolor{armygreen}{rgb}{0.29, 0.33, 0.13}
\definecolor{asparagus}{rgb}{0.53, 0.66, 0.42}

%\textcolor{blue}
%\textcolor{asparagus}
\newcommand{\slow}{{\mathfrak{b}}}
\newcommand{\shigh}{{\mathfrak{g}}}

\title{Rational Groupthink}
\author{Matan Harel$^1$}
\address{$^1$Tel Aviv University}
\author{Elchanan Mossel$^2$}
\address{$^2$Massachusetts Institute of Technology }
\author{Philipp Strack$^3$}
\address{$^3$Yale University}
\author{Omer Tamuz$^4$}
\address{$^4$California Institute of Technology}
\begin{abstract}
We study how long-lived rational agents learn from repeatedly observing a private signal and
each others' actions.
%We find that in the long run, information aggregation
%fails, and the fraction of private information transmitted goes to
%zero as the number of agents gets large.
With normal signals, a group of any size learns more slowly than just four agents who directly observe each others' private signals in each period. Similar results apply to general signal structures. We identify \emph{rational} \textit{groupthink}\textemdash in
which agents ignore their private signals and choose the same action
for long periods of time\textemdash as the cause of this failure of
information aggregation.

\medskip

{\bf JEL}: D83.

\thanks{We thank seminar audiences in Berkeley, Berlin, Bonn, Caltech, Chicago,
Düsseldorf, Duke, Harvard, Medell\'in, Microsoft Research New England,  MIT, Montreal,
NYU, Penn State, Pittsburgh, Princeton, San Diego, UPenn, USC and
Washington University, Yale, as well as Nageeb Ali, Ben Brooks, Dirk Bergemann,
Kim Border, Federico Echenique, Wade Hann-Caruthers, Benjamin Golub,
Rainie Heck, Paul Heidhues, Shachar Kariv, Navin Kartik, Steven Morris,
Luciano Pomatto, Larry Samuelson, Lones Smith, Juuso Toikka, Leeat
Yariv and others for insightful comments and discussions. Matan Harel was partially supported by the IDEX grant of
Paris-Saclay. Elchanan
Mossel is supported by ONR grant N00014-16-1-2227 and NSF grant CCF
1320105. Omer Tamuz was supported by a grant from the Simons Foundation (\#419427). }
\end{abstract}

\maketitle

\section{Introduction}
A key question in social learning is: How well do agents learn by observing each others' actions?
As the analysis of the beliefs of long-lived Bayesian
agents is challenging\footnote{See e.g.\ \cite*{cripps2008common}.}, most of the literature focuses either on short-lived agents\footnote{See e.g.\ \cite*{dasaratha2018social,mueller2015multi}.}
or on non-rational belief dynamics such as the DeGroot model or quasi-Bayesian agents\footnote{See \cite*{golub2010naive,molavi2015foundations}.}. 
By applying large deviation techniques we
overcome the difficulty associated with the analysis of Bayesian beliefs and analyze social learning with long-lived rational agents.

Formally, we consider a group of myopic Bayesian agents who repeatedly observe private signals about a binary state, as well as each others' past actions.
As every agent eventually learns the state, our main focus is on the speed of learning, i.e.\ the rate with which the agents converge to the correct action.
Our main result is that information aggregation will fail for Bayesian agents
in a large society: An arbitrarily large group of Bayesian agents
observing each others' actions will only learn as fast as a small
group of agents observing each others' signals directly. For example,
when signals are normal, four agents sharing their signals learn faster
than a group of arbitrarily many agents who observe each others' actions, but not signals. 

This failure of information aggregation is caused by endogenous correlation in the agents' actions, which reduces the amount of information that these actions reveal about the private signals.\footnote{As is well known,
a large number of sufficiently correlated signals convey less information
than a small number of independent signals.}
 Whereas signals are independent, the agents' actions become correlated, as they are all heavily influenced by the past, observed actions. This correlation is an immediate consequence of the incentive
to learn from each others' actions. For example, if agent 1 takes
an action that is optimal in some state of the world, the other agents
will infer that agent 1's private belief indicates that this state
is relatively likely and will themselves take this action with greater
probability. A greater number of agents increases this correlation, as agents share more common information. This decreases the amount of information that the agents' actions reveal about their private signals. The insight of our analysis is that as the number of agents grows, the
correlation increases to an extent that completely out-weighs the
gain of the additional independent private signals. We show that asymptotically
this failure of information aggregation holds for any signal structure,
any utility function and any number of agents.

What inference an agent draws from the actions of another agent depends
on her belief about the other agent's belief. Thus, agents' actions
may depend on their higher order beliefs. This poses a significant
challenge for the exact characterization of behavior. We circumvent
this problem by focusing on long-term probabilities, and
by analyzing a phenomenon that we call ``rational groupthink'', conditional on which higher-order beliefs admit a tractable structure. We define rational groupthink to be the event that \textit{all} agents take the wrong action for many periods. We show that when this event occurs, it is likely that all agents have
private signals that indicate the correct action. Through a fixed-point
argument we are able to estimate the asymptotic probability of rational
groupthink, and
find that rational groupthink occurs so often that agents in a large group learn
almost as slowly as they do in autarky. Hence, in this sense, rational
groupthink prevents almost all information aggregation.\footnote{Our prediction seems to be in line with the findings in the empirical
literature: \citet*[page 5]{da2016harnessing} find in a study on forecasters
``that private information may be discarded when a user places weights
on the prior forecasts {[}of others{]}. In particular, errors in earlier
forecasts are more likely to persist and appear in the final consensus
forecast, making it less efficient.''}

Groupthink arises in our model as a consequence of Bayesian updating, rather than being driven by an assumed desire for conformity.\footnote{See \cite*{angeletos2007efficient} for a setting in which payoff externalities lead to a desire for conformity, which in turn lead agents to discount their private signals.} Rational groupthink occurs after a consensus on an action is formed
in the initial periods, making it optimal for every agent to continue
taking the consensus action, even when her private information indicates
otherwise. Indeed, we show that typically, after a wrong consensus
forms, all agents eventually observe private signals which provide strong
evidence for choosing the correct action, and yet a long time may
pass until any of them breaks the wrong consensus (Theorem~\ref{prop:groupthink}).
Thus a situation arises in which each agent's private information
indicates the correct action, and yet, because of the group dynamics,
all agents choose the wrong action. 

We study the effect of increasing the group size. On the one hand,
with more agents, each individual agent is less likely to break a
wrong consensus. On the other hand, the number of potential dissenters
is larger, and so a priori it is not obvious whether rational groupthink
becomes more or less likely. 
%We show that the  share of information that is lost due to the rational groupthink effect becomes arbitrarily large as the size of the group increases. 
Our first main result shows that, even as the number
of agents goes to infinity, the speed of learning from actions stays
bounded by a constant (Theorem \ref{thm:speed-bound-1}), whereas
the speed of learning from the aggregated signals, which is proportional
to the number of agents, goes to infinity (Fact \ref{fact:public-signals}).
Thus, in a large group, almost no information is aggregated; the agents'
beliefs when observing only actions have the same precision as would
result from observing a vanishingly small fraction of the available
private signals. Specifically, for normal signals, a group of $n$
agents observing each others' actions learns asymptotically slower
than a group of $4$ agents who share their private signals; this
holds for any number of agents! Hence, at most a fraction of $\nicefrac{4}{n}$
of the private information is transmitted through actions (Corollary
\ref{cor:Inefficiency}). We proceed beyond normal signals to show
that for any signal distribution at most a fraction of $\nicefrac{c}{n}$
of the private information is transmitted through actions, for some
constant $c$ that depends only on the distribution of the
private signals (Lemma~\ref{claim:q-bound}).

As mentioned above, we quantify the speed of learning as the asymptotic rate at which agents converge to the correct action. An advantage of asymptotic rates is that they are independent of many details of the model, providing a measure that is robust to changes in model parameters such as the agents' prior or the exact utility function. Furthermore, they are tractable. For similar reasons of tractability and robustness, many previous works have studied asymptotic (long run) rates of learning in various settings.\footnote{Examples of papers studying the rate of learning are \citet*{vives1993fast,chamley2004rational,duffie2007information,duffie2009information,duffie2010information}. Asymptotic rates also have been studied in other settings in which
it is difficult to analyze the short-term dynamics \citep*[e.g.,][]{hong2004rates,horner2016fast}.
\citet{molavi2015foundations}
study the rate of learning in an almost identical setting, with boundedly
rational agents.}

As a robustness test, we complement our results on the asymptotic rates by an analysis of the probability with which the wrong action is chosen in early periods. We study a canonical setting of a large group of agents with normal private signals, where, as the size of the group is increased, the total precision of their aggregate signal is kept constant. This regime guarantees that the total amount of information available to society is independent of the number of agents, allowing us to study groups which can be very large, but still not learn immediately. 

Using numerical simulations, we show that, for example, the probability with which an agent makes a mistake in period 10 equals roughly 14\% with either 40 or 100 agents observing each others' actions, but equals 0.07\% and is thus 200 times smaller, if signals are public. These and other simulation results indicate that  our asymptotic results---which only have formal implications for late enough periods---sometimes already hold in the early periods.

We complement these simulations with our second main theorem, which shows that in this setting, as the number of agents goes to infinity, the probability that an agent chooses correctly in some given period tends to the (roughly constant) probability with which the majority of agents choose correctly in the first period  (Theorem \ref{thm:short-term}). This is because, after observing the first period actions, agents will tend to ignore their private signals for many periods. Thus, when the group is large, the private signals of period two and later periods are effectively lost, and information fails to aggregate not only asymptotically, but already after the first period.

A defining feature of our model is that information flows bidirectionally. In \S\ref{sec:incomplete} we study a setting in which information flows only in one direction, and show that there, a non-vanishing fraction of information is aggregated. We consider a partial observation structure in which agent 1 observes the actions of all others in addition to his private signals, and each of the remaining agents observes his private signals only. In this setting, agent 1 will learn with a speed that increases linearly with the number of agents (Theorem~\ref{prop:observing-actions}), in sharp contrast to our main result. 
This highlights that the almost complete failure of information aggregation in our baseline setting occurs because of the bidirectional flow of information, and not just because agents observe actions rather than signals. This is in contrast to the herding literature where information aggregation fails even when information travels only unidirectionally.

We assume throughout that agents are Bayesian and myopic: they completely discount future payoffs, and thus at every time period choose the action that maximizes the expected payoff at that period. In a repeated action setting with non-myopic agents there may be a strategic incentive to change ones own action in order to gain more information from future actions of others. This effect does not exist for rational myopic agents, and we make this assumption for tractability, as does most of the learning literature.\footnote{Indeed, the same choice is made in most of the learning literature (where signals are private and agents interact repeatedly) either explicitly \citep[e.g.,][]{sebenius1983don,parikh1990communication,bala1998learning,keppo2008optimal}, or implicitly, by assuming that there is a continuum of agents \citep[e.g.,][]{vives1993fast,gale2003bayesian,duffie2007information,duffie2009information,duffie2010information}.} A possible justification for this approach is that reasoning about
the informational effect of one's actions in such setups requires
a level of sophistication that seems unrealistic in many applications.\footnote{We conjecture that all our results generalize to the case of non-myopic
agents, but this extension requires substantial technical innovation,
beyond the techniques developed in this paper.}

\subsection*{Related Literature}
Most of the preceding literature studies situations where each agent
observes a \textit{single} signal and agents try to infer the others'
signals from repeatedly observing their actions. \citet*{geanakoplos1982we,sebenius1983don,parikh1990communication,mossel2012strategic}
give conditions under which agents actions agree in the long-run.
\citet*{rosenberg2009informational} show that, for a large class of social learning models, agents reach asymptotic agreement, provided the agents observe enough information about each others' actions. The question of how well information
is aggregated in such settings was considered in an important paper
by \citet{vives1993fast}, who studies the rate at which information is
aggregated through noisy prices. 

In contrast to this literature we allow for agents to repeatedly observe
signals about the state of the world and the actions of others. The only other article which
we are aware of that tackles this  problem is \citet{molavi2015foundations}, which studies asymptotic rates of
learning under (non-rational) linear belief updating rules in complex
observational networks. The focus of this paper differs from ours: they
allow for complex network structures, but impose simple linear belief
updating rules. In contrast, we study the complexities associated
with Bayesian learning, but assume that all actions are commonly known.
Interestingly, our results contrast their findings; while in their model information
is quickly aggregated, in our model it is not. This is in part a consequence
of the difference in the rationality assumptions.\footnote{In \S\ref{sec:non-bayesian} we present numerical simulations that indicate that non-Bayesian updating could lead to faster learning in our setting.}

\citet{gale2003bayesian} use numerical methods to characterize the
asymptotic rates with which rational agents learn, and emphasize the
importance of understanding the rates at which Bayesian agents learn
from each other.\footnote{\citet[p.20]{gale2003bayesian}: ``Speeds of convergence can be established
analytically in simple cases. For more complex cases, we have been
forced to use numerical methods. The computational difficulty of solving
the model is massive even in the case of three persons {[}...{]} This
is an important subject for future research.'' }

Our work is also related to models of rational herding
as we use the same conditional i.i.d.~structure of signals, and utilities
depend only on one's own actions and the state.\footnote{See the original papers of \cite*{bikhchandani1992theory} and \cite*{banerjee1992simple}, as well as  \cite*{smith2000pathological, chamley2004rational, acemoglu2011bayesian, rosenberg2017efficiency}, and many others.} While in sequential models the number of agents is equal to the number of time periods, in our model these can be varied independently.\footnote{As an exception, \cite{dasaratha2019speed} recently study the effect of changing population size in a sequential model with overlapping generations.}

A more significant difference is that in herding models each agent acts only once, and thus information is transmitted between agents only in one direction, which implies that higher order beliefs to play no role. A contribution of this paper is to show that the failure
of information aggregation is not particular to sequential models, but more generally extends to situations
of repeated interactions. Our main finding, the rational groupthink
effect, has no analogue in sequential herding models, since, in these
models, once a herd starts, it is not true that every agent's private
signal indicates the correct action.

Our work is also related to the literature that studies how groupthink arises from various other motives. \cite*{benabou2012groupthink} shows that anticipatory or Kreps-Porteus preferences can lead agents to willingly ignore freely available information if others chose to ignore it even absent any social learning. \cite*{ottaviani2001information} demonstrate how a desire to appear well informed can lead to herding and groupthink. \cite{angeletos2007efficient} show how a desire to coordinate actions can lead agents to discount their private signals in favor of public information.

Potential applications of our results appear in settings in which
agents repeatedly learn from each other. These include the dissemination
of information in developing countries (e.g., \citet{conley2010learning,banerjee2013diffusion}
among many studies), the adoption of opinions on social networks,
and prediction markets where forecasters observe the forecasts of
others \citep[see][]{da2016harnessing}.

\section{Leading Example: Aggregating Information Through Prices}\label{sec:leading-example}

As a leading economic example, consider local monopolistic sellers who want to learn about the quality of a new product and the associated optimal price. For concreteness, imagine that each seller is the owner of a theater / shop in a different city, and has to decide how much to charge for a new movie, musical, book, toy or fashion item. Because sellers act in different markets there are no direct payoff externalities. Assume that the product is either good or bad, with corresponding demand either high or low. As the demand in other markets is informative about the product's quality, it is also informative about demand in the seller's home market. When marginal profits are not constant in the volume of sales, a seller will want to set one price if the demand is high, another price if the demand is low, and potentially intermediate prices when she is unsure about the state. Consequently, each seller wants to learn the state and can do so not only by observing her local demand, but also by observing the prices set by other sellers.

A second application is that of learning about the quality of a new government policy (e.g., Obamacare) via social media. A group of people who differ in their location, income, family status, etc., each receive private signals from their experience with the new policy, and share their coarse opinion of it on social media, where they also observe the opinions of others.\footnote{A less economic example---which we nevertheless find compelling---is that of a group of agents who are interested in knowing whether or not there is a god. Every morning, each toasts bread for breakfast, and checks to see if a divine signal appears in the burn patterns. If there is no god, then the probability of this event is very low. If there is a god, then this probability is significantly higher, although still low. After breakfast, people declare to each other whether or not they believe in god, based on some common threshold of belief. The importance of minor miracles to the belief in god has a long history; for example, in Judaism and Christianity, the concept of {\em special providence} (Hebrew: {\em hashgacha pratit}, literally meaning ``private monitoring'') refers to the idea that god frequently performs small miracles to his believers \citep[e.g.,][part 3, chapters 17-18]{Maimonides}.  \cite{hume1748ofmiracles} discusses the reliability of reports of miracles and their implications on beliefs. See  \cite{holder1998hume} for a modern discussion of Bayesian updating of the belief in god following the reporting of miracles.  For a recent example see, e.g., \url{http://news.bbc.co.uk/2/hi/americas/4019295.stm}.  We thank the editor for suggesting this example. }

To model these situations, we assume that each seller decides each period which price to charge, and then observes demand in her city, as well as the prices charged by sellers in other cities. To illustrate our main results in the cleanest possible setting, we make a number of simplifying assumptions in this section: 
\begin{compactenum}
\item[(i)] the sellers choose between only two prices;
\item[(ii)] each seller receives a payoff of $1$ if she charges the high price and the product is good or she charges a low price and the product is bad, and a payoff of 0 otherwise;
\item[(iii)] the signal the seller observes at each period is normally distributed with mean $-1$  when demand is low, and mean $+1$ when demand is high;
\item[(iv)] Finally, the variance of each seller's signal equals $n$; this ensures that the total information contained in all sellers combined signals is constant and allows us to compare outcomes for different numbers of sellers.\footnote{The analogue of this assumption in a setting with Binary signals would be to keep the total number of signals fixed and have each agent privately observe an equal share of these signals.}
\end{compactenum}
We consider  more than two possible prices, arbitrary payoff functions, and arbitrary signal distributions in our general results which we present in \S\ref{sec:setup} and the following sections.\footnote{For a more realistic model of random demand within our framework, one could assume that the number of customers interested in buying the product is Poisson distributed, with parameter depending on the state.}

\subsection{Speed of Learning from Actions}
For this setup we used Monte Carlo simulations to compute the error probabilities in each period, as well as other quantities of interest (see \S\ref{app:simulations} for details); we use the term ``error'' to describe the choice of an action that is not optimal, given the realized state, such as choosing the high price when the product is bad.
\begin{figure}[t]
  \includegraphics[width=\linewidth]{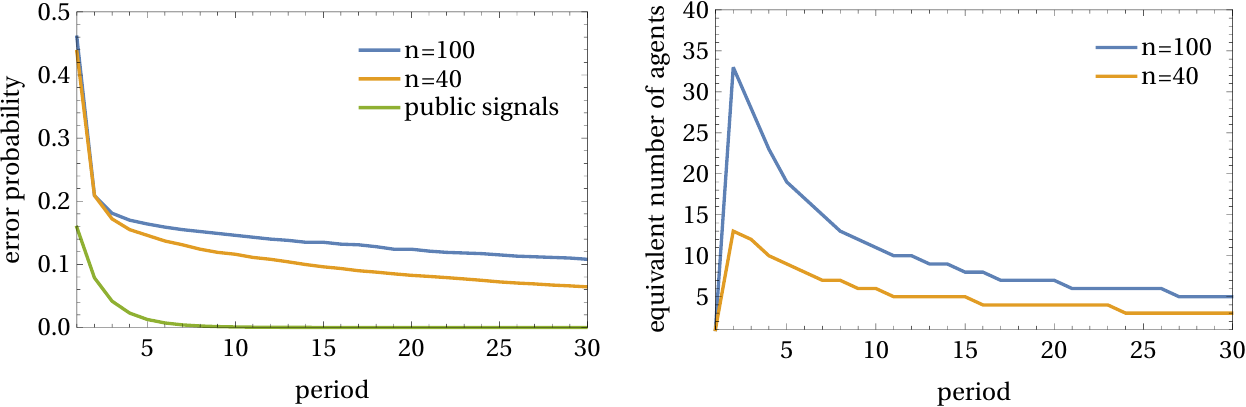}
  \caption{The probability with which an agent takes the wrong action over time in absolute terms (on the left) and the number of agents sharing their signals which would lead to the same probability of error (on the right).\label{fig:errorprobabilities}}
  
\end{figure}
Figure \ref{fig:errorprobabilities} displays the results of these simulations. The plot on the left side shows the error probability for different group sizes when observing actions, and for the case where all signals are public. What immediately stands out is how much slower a group of agents learns by observing each others' actions relative to observing each others' signals. For example, the probability with which an agent makes a mistake in period 10 equals roughly $14\%$ with either 40 or 100 agents observing each others' actions, but equals $0.07\%$ and is thus 200 times smaller, if signals are public; note that by our choice of variance the probability of error with public signals is independent of the number of agents. 

Another way one could measure how much information is lost due to the fact that agents observe actions is by looking at the smallest number of agents that could match the probability of error by sharing their signals. We draw this illustration in the right plot of Figure~\ref{fig:errorprobabilities}. For example, in period 30 we have the following comparisons: 3 agents sharing their signals are less likely to take the wrong action than 40 agents observing each others' actions. And 5 agents sharing their signals are less likely to take the wrong action than 100 agents observing each others' actions. Thus, in this example $92\%$ (resp., $95\%$) of the information contained in the agents' private signals is lost when the agents learn only from actions. We establish in Theorem \ref{thm:speed-bound-1} that this phenomenon is not due to the specific set of parameters chosen in the example: whenever signals are normal, 4 agents sharing their signals are eventually more likely to make the correct choice than $n$ agents observing each others' actions, for any $n$! For non-normal signals the same result holds, with the number $4$ replaced by a constant depending on the distribution, and given explicitly  in Theorem \ref{thm:speed-bound-1}.

\subsection{Why Learning from Actions is Slow}
\begin{figure}[t]
  \includegraphics[width=\linewidth]{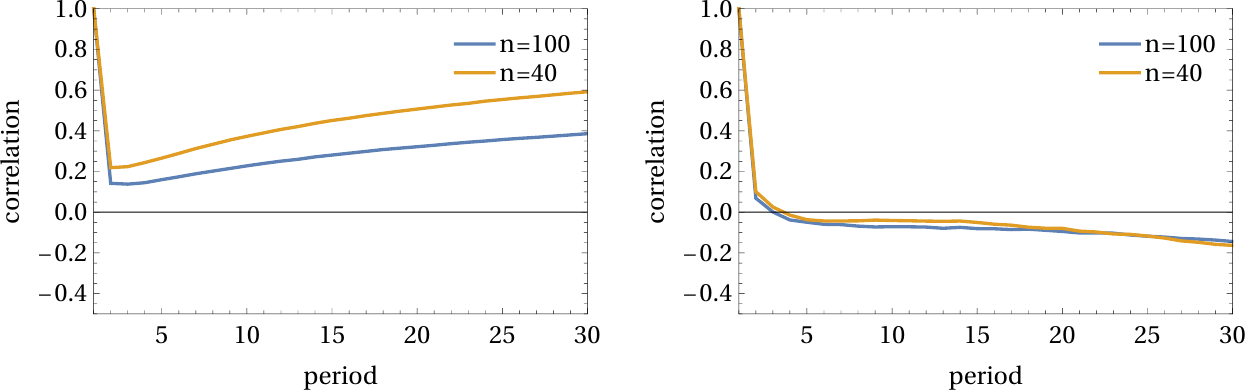}
  \caption{Correlation between the action taken by an agent, and the action the agent would take based on her private signals only (on the left). On the right is the correlation, conditioned on the agent choosing the wrong action.%The probability with which and agent has private signals indicating the correct action conditioned on the agent choosing the wrong action (on the right).
  \label{fig:correlation}
  }
  
\end{figure}

Why is it that most information is lost when only actions are observed? To better understand this phenomenon it is instructive to study the correlation between an agent's actions and his private signals. We plot this correlation on the left in Figure~\ref{fig:correlation}.
This correlation is a (rough) measure of the information that can be inferred from an agent's actions. In the first period each agent does not observe any information from other agents and thus chooses her action based solely on her first period signal. This leads to a correlation of 1 and the revelation of all private information in the first period. As agents' first period signals are independent conditional on the state, so are their actions. Thus, when observing the others' first period actions, an agent observes many conditionally independent signals.\footnote{As a consequence, the probability of error drops considerably between the first and second period; see Figure~\ref{fig:errorprobabilities}.}
When the number of agents is large the information in the first period actions is likely to lead to a stronger signal than each agent's private signals in subsequent periods. As a consequence, agents are likely to follow the action taken by the majority in the first period.  This in turn leads agents  to not condition their actions on their own signals, making future actions uninformative and hindering information aggregation. The sudden drop in correlation between the agent's action and private signals in the second period shown in Figure~\ref{fig:correlation} illustrates this effect. It is apparent from Figure~\ref{fig:correlation} that the low correlation between the agents' actions and private signals prevails for many periods, and is significantly lower for 100 compared to 40 agents. This is formalized in Theorem~\ref{thm:short-term}, which shows that given a sufficiently large number of agents, in any given period after the second, all agents with high probability ignore their private signals, leading to a small correlation between actions and private signals.

\subsection{Groupthink}
\begin{figure}[t]
  \includegraphics[width=\linewidth]{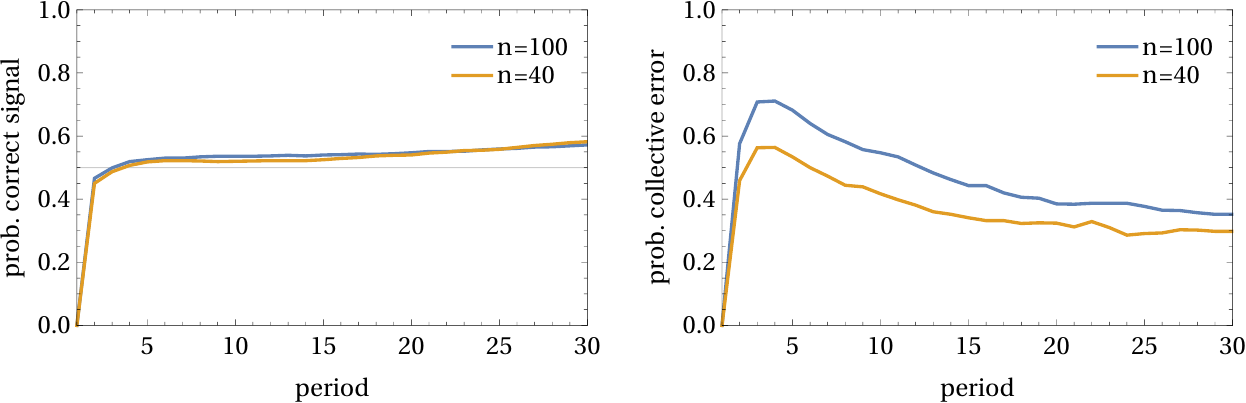}
  \caption{Probability of correct private signals (on the left) and all agents taking the wrong action (on the right) conditional on taking the wrong action in a period.\label{fig:correlation2}
  }
  
\end{figure}

The right plot of Figure~\ref{fig:correlation2} shows that the event in which all agents choose incorrectly does not have insignificant probability, but in fact happens often, conditional on an agent choosing incorrectly. Thus, a signal agent making a mistake is closely tied with the entire group making a mistake.

The right plot of Figure~\ref{fig:correlation} shows that when the agents choose the incorrect action, their private signals are negatively correlated with their actions. In other words, conditioned on making a mistake, an agent is likely to have a correct private signal. For example, as is shown on the left side of Fig.~\ref{fig:correlation2}, conditioned on choosing the wrong action, an agent's private signal indicates the correct action with probability 57\%. This may be surprising, as one might have reasonably expected that when an agent chooses the wrong action, it is because of incorrect private signals.

This is formalized in Theorem~\ref{prop:groupthink}, which captures the groupthink effect: agents take the incorrect action because of the group influence, and despite having the correct signal. Moreover, no agent needs to have an incorrect private signal for all agents to choose the incorrect action. In fact, as Theorem~\ref{prop:groupthink} shows, when all agents take the wrong action in late periods, they all, with high probability, have private signals indicating the correct action.

%

%
% Discussion of correlation
%
\section{Setup}\label{sec:setup}

Time is discrete and indexed by $t\in\{1,2,\ldots\}$. Each period,
each agent $i\in\{1,2,\ldots,n\}$ first observes a signal (or shock)
$s_{t}^{i}\in\mathbb{R}$, takes an action $a_{t}^{i}\in A$, and
finally observes the actions taken by others this period. The set
of possible actions is finite: $|A|<\infty.$

\subsection{States and Signals}

There is an unknown state 
\[
\Theta\in\{\slow,\shigh\}
\]
randomly chosen by nature, with probability $p_{0}=\mathbb{P}\left[\Theta=\shigh\right]\in(0,1)$.
For ease of exposition, we call $\slow$ the bad state and $\shigh$ the good state, even though the model is completely symmetric in the state.
Signals $s_{t}^{i}$ are i.i.d, across agents $i$ and over time $t$,
conditional on the state $\Theta$, with distribution $\mu_{\Theta}$. Throughout, we denote by $\mathbb{E}_{\theta}\left[\cdot\right]:=\mathbb{E}\left[\cdot\mid\Theta=\theta\right]$
and $\mathbb{P}_{\theta}\left[\cdot\right]:=\mathbb{P}\left[\cdot\mid\Theta=\theta\right]$
the expectation and probability conditional on the state.
The distributions $\mu_{\shigh}$ and $\mu_{\slow}$ are mutually absolutely
continuous\footnote{That is, every event with positive probability under one measure has
positive probability under the other.} and hence no signal perfectly reveals the state. As a consequence
the log-likelihood ratio of every signal 
\[
\ell_{t}^{i}=\log\frac{\mathrm{d}\mu_{\shigh}}{\mathrm{d}\mu_{\slow}}(s_{t}^{i})
\]
is well defined (i.e., $|\ell_{t}^{i}|<\infty$) and we assume that
it has finite expectation $|\mathbb{E}\left[\ell_{t}^{i}\right]|<\infty$.
We also assume that priors are generic\footnote{That is, chosen from a Lebesgue measure one subset of $[0,1]$.},
so as to avoid the expository overhead of treating cases in which
the agents are indifferent between actions; the results all hold even
without this assumption.

Our signal structure allows for bounded as well as unbounded likelihoods.\footnote{In the herding literature agents either learn or do not learn the
state, depending on whether private signals have bounded likelihood
ratios \citep{smith2000pathological}. In our model, the distinction
between unbounded and bounded private signals is not important, since
the aggregate of each agent's private information suffices to learn
the state.} Our main example is that of \textit{normal signals} $s_{t}^{i}\sim\mathcal{N}(m_{\theta},\sigma^{2})$
with mean $m_{\theta}$ depending on the state and variance $\sigma^{2}$.
Another example is that of \textit{binary signals} $s_{t}^{i}\in\{\slow,\shigh\}$
which are equal to the state with constant probability $\mathbb{P}_\theta\left[s_{t}^{i}=\theta\right]=\phi>\nicefrac{1}{2}$.

\subsection{Actions and Payoffs}

Agent $i$'s payoff (or utility) in period $t$ depends on her action
$a_{t}^{i}$ and next period's signal $s_{t+1}^{i}$, and is given
by $u(s_{t+1}^{i},a_{t}^{i})\,.$\footnote{Note, that observing the utility $u(s_{t+1}^{i},a_{t}^{i})$ does
not provide any information beyond the signal $s_{t+1}^{i}$ and therefore
past signals $(s_{1}^{i},\ldots,s_{t+1}^{i})$ are a sufficient statistic
for the private information available to agent $i$ when taking an
action in period $t+1.$}
%The signal can be interpreted as a shock (like demand or interest rate) which influences the payoffs of the different actions of the agent. 
Note that $u(\cdot,\cdot)$ does not depend on the agent's
identity $i$ or the time period $t$.\footnote{This model is equivalent to
a model where the agent's utility $\bar{u}(\Theta,\,a_{t}^{i})$ is
unobserved and depends directly on the state. Formally, we can translate
the model where the utility depends on the signal into the model where
it depends on the state by setting it equal to the expected payoff
conditional on the state $\theta \in \{\slow,\shigh\}$ 
%\begin{align*}
$\bar{u}(\theta,\act) :=\mathbb{E}_{\theta}\left[u(s_{t+1}^{i},\act)\right]\,$. }
We denote by $a^{\theta}$ the action that maximizes the flow payoff
in state $\theta$, which we assume is unique 
\[
\act^{\theta}:=\arg\max_{\act\in A}\,\bar{u}(\theta,\act)\,.
\]
We call $\act^{\shigh},\act^{\slow}$ the \textit{certainty actions} and assume
that they are distinct (i.e., $\act^{\shigh}\neq\act^{\slow}$), as otherwise
the problem is trivial.

It is an important feature of this model that externalities are purely
informational, i.e., each agent's utility is independent of the others'
actions, and hence agents care about others' actions only because
they may provide information. Furthermore, private signals are independent
of actions, and so agents have no experimentation motive; they learn
the same information from their signals, irregardless of the actions
that they take. 

\subsection{Agents' Behavior and Information}

We denote by $p_{t}^{i}$ the posterior probability that agent $i$
assigns to the event $\Theta=\shigh$ after observing her private signal and before choosing her period $t$ action.

We assume throughout that agents are myopic: at each period they choose the action that maximizes their stage utility, completely discounting the future. As an agent's posterior belief $p_{t}^{i}$ is a sufficient statistic
for her expected payoff, her action $a_{t}^{i}$ almost surely depends only on $p_{t}^{i}$.\footnote{This statement holds, as for almost every prior the agent will not be indifferent.} 
%Formally, there exists a function $a^{\star}\colon[0,1]\to A$ such
%that with probability one
%\[
%a_{t}^{i}=a^{\star}(p_{t}^{i})\,.
%\]
%there exists intervals $(\underline{p}_i,\overline{p})$

%As information arrives independently of actions, and because agents are myopic, our model is not one of strategic experimentation: there are no incentives to change one's action in order to learn more from one's own future signals, or from others' future actions. With these potentially confounding effect removed, we are left with a distilled model that allows us to study how observing others' actions differs from observing their signals.

Each agent observes only her own signals, and not the signals of others.
To learn about the state, agents try to infer the signals of others
from their actions. More precisely, at the end of each period an agent
observes the actions taken by all other agents in this period. 

%\subsection{Examples}

\subsubsection{\label{subsec:Matching-the-State}Example: Matching the State}

Our model allows for any (finite) number of actions, and any signal distribution. Nevertheless, a simple example which suffices to understand all the economic results
of the paper is the case of two actions $A=\{\slow,\shigh\}$ where the agent's
expected utility equals one if she matches the state, i.e. 
\[
\bar{u}(\theta,\act)=\begin{cases}
1 & \text{ if }\act=\theta\\
0 & \text{ if }\act\neq\theta
\end{cases}\,.
\]
In this case the agent simply takes the action to which her posterior
belief assigns higher probability: 
\[
a_{t}^{i}=\begin{cases}
\shigh & \text{ if }p_{t}^{i}>\frac{1}{2}\\
\slow & \text{ otherwise }
\end{cases}\,.
\]

%\subsubsection{Monopolistic Sellers}\label{subsec:MonoplisticSellers}

%As an application, consider local monopolistic sellers who want to learn about the demand for their product and the associated optimal price. Each seller acts in a different market, so that there are no payoff externalities. The distribution of demand, however, is the same, so that the realized demand in other markets is informative about future demands in a seller's home market.

%For concreteness, assume that the sellers are shop owners who are selling a new product, and that in the good state the number of people entering the store to inquire about the product is Poisson with mean $\rho_{\shigh}$, while in the bad state it is Poisson with mean $\rho_{\slow}$, which is less than $\rho_{\shigh}$. After learning the price each customer decides whether or not to buy, depending on her private valuation. Customers' private valuations for the product are independent of the state, and so, after having entered the store, customers reveal no new information about the state. Thus, the information a seller learns about the state from her own customers is independent of the price she sets.

%When marginal profits are not constant in the volume of sales, a seller will want to set one price if the state is good, another price if the state is bad, and potentially intermediate prices when she is unsure about the state. Consequently, each seller wants to learn the state and does so not only by observing the demand in her store, but also by observing the prices set by other sellers.

\section{Results}

\label{sec:result}In this section we describe our results; \S\ref{sec:Learning-Dynamics} derives the learning dynamics in detail
and explains how they lead to the results of this section. We consider
the probability with which an agent $i$ takes a suboptimal action
in period $t$: 
\[
a_{t}^{i}\neq\alpha^{\Theta}\,.
\]
We refer to this event as agent $i$ ``making a mistake'' by ``choosing
the wrong action'', even though she takes the action which is optimal
given her information. As a benchmark we first briefly discuss the
classical single agent case.

\subsection{Autarky}\label{subsec:results-autarky}

In the single agent case $n=1$, the probability of a suboptimal action
is known to decay exponentially, with a rate $\ra$ that can be calculated
explicitly in terms of the cumulant generating functions\footnote{Here $\ell$ is a random variable with a distribution that is equal
to that of any of the log-likelihood ratios $\ell_{t}^{i}$.} $\lambda_{\shigh}(z):=-\log\,\mathbb{E}_{\shigh}\left[\mathrm{e}^{-z\,\ell}\right]$
and $\lambda_{\slow}(z):=-\log\,\mathbb{E}_{\slow}\left[\mathrm{e}^{z\,\ell}\right]$:\footnote{The signs used in this definition deviate from the standard definition $\log\mathbb{E}_{\theta}[\mathrm{e}^{z\ell}]$ of the cumulant generating function of $\ell$. Our choice allows for a convenient formulation of Lemma~\ref{lem:LLRDeviations} below, and reflects the fact that in the good state, high $\ell$ indicates a correct signal, while in the low state it indicates an incorrect one.}
\begin{fact}[Speed of learning in autarky]
\label{thm:ErrorProbSingleAgent-1}The probability that a single
agent in autarky chooses the wrong action in period $t$ satisfies\footnote{Here, and elsewhere, we write $o(t)$ to mean a lower order term.
Formally a function $f\colon\mathbb{R}\to\mathbb{R}$ is in $o(t)$
if $\lim_{t\to\infty}f(t)/t=0$. } 
\begin{equation}
\mathbb{P}\left[a_{t}\neq\act^{\Theta}\right]=\mathrm{e}^{-\ra\cdot t+o(t)}\,,\label{eq:RatesSingleAgent-1}
\end{equation}
where
\[
\ra:=\sup_{z\geq0}\lambda_{\shigh}(z)=\sup_{z\geq0}\lambda_{\slow}(z).
\]
\end{fact}
We provide a formal proof of this fact in the appendix, where we also explain why $\sup_{z\geq0}\lambda_{\shigh}(z)=\sup_{z\geq0}\lambda_{\slow}(z)$.
This type of autarky result is classical in the statistics literature
and can be found, for example, in studies of Bayesian hypothesis testing;
see, e.g.~\citet[pages 314-316]{cover2012elements}. For us it serves
as a benchmark for the case when agents try to learn from the actions
of others. We prove Fact \ref{thm:ErrorProbSingleAgent-1} in the
Appendix, for the convenience of the reader.

Note, that the long-run probability of a mistake is independent of the
set of actions and the utility function. It is also independent
of the prior. Thus quantifying the speed of learning using the exponential
rate has both advantages and disadvantages: the rate is independent
of many details of the model and depends only on the private signal
distributions. It is also tractable and can be explicitly calculated
for many distributions. However, it is an asymptotic measure and in
general does not say anything formally about what happens in early
periods.
\subsection{Many agents}

We now turn to the case where there are $n\geq2$ agents. We first
consider the benchmark case where all signals are observed by all
agents. Since there is no private information, all agents hold the
same beliefs, and this case reduces to the single agent case, but
where $n$ signals are observed in every period. After $t$ periods
the agents will have observed $n\cdot t$ signals, and so, by Fact
\ref{thm:ErrorProbSingleAgent-1}, their probability of taking the
wrong action will be the probability of error after $n\cdot t$ periods
in the autarky setting.
\begin{fact}[Speed of learning with public signals]
\label{fact:public-signals}When signals are public, the probability
that any agent $i$ chooses the wrong action in period $t$ satisfies
\end{fact}
\[
\mathbb{P}\left[a_{t}^{i}\neq\act^{\Theta}\right]=\mathrm{e}^{-n\,\ra\cdot t+o(t)}.
\]

Having considered this benchmark case, we turn to our model,
in which $n\geq2$ agents observe each others' actions, but signals
are private. Our main result is that for any number of agents the
speed of learning is bounded from above by a constant: 
\begin{thm}
\label{thm:speed-bound-1}Suppose $n$ agents all observe each others'
past actions. Given the private signal distributions, there exists
a constant $\rmax>0$ independent of the number of agents $n$, such that 
\[
\mathbb{P}\left[a_{t}^{i}\neq\alpha^{\Theta}\right]\geq\mathrm{e}^{-\rmax\cdot t+o(t)}.
\]
In particular, this holds for $\rmax=\min\left\{ \mathbb{E}_{\shigh}\left[\ell\right],-\mathbb{E}_{\slow}\left[\ell\right]\right\} $.
When private signals are normal then one can take $\rmax=4\,\ra$.
\end{thm}
Note that this theorem holds for all fixed signal distributions and all group sizes $n$, and does not require any assumptions about the relation between them, such as the ones we make in \S\ref{sec:leading-example}.

An immediate corollary from Theorem \ref{thm:speed-bound-1} and Fact
\ref{fact:public-signals} is the following result.
\begin{cor}
\label{cor:Inefficiency}There exists a fixed group size $k$ such
that for any arbitrarily large group size $n$, the probability that
any agent chooses the wrong action is eventually lower with $k$ agents
and public signals than with $n$ agents who only observe actions. When
signals are normal we can take $k=4$.
\end{cor}
Thus, adding more agents (and with them more private signals and more
information) cannot boost the speed of learning past some bound, and
as $n$ tends to infinity more and more of the information is lost: a vanishing fraction of the private signals would produce the  the same error probabilities if observed directly.\footnote{Formally, Theorem \ref{thm:speed-bound-1} establishes that the upper bound on the rate of learning, $\rmax$, is less than some constant times $1/n$ times the rate $n \ra$ of learning from observing $n$ signals directly every period, i.e. $\frac{\rmax}{n \ra} \leq \frac{c}{n}$, and thus goes to zero for $n$ tending to $\infty$.}
In the case of normal signals $\rmax=4\,\ra$, and thus, regardless
of the number of agents, the probability of mistake is eventually
higher than it would be if $4$ agents shared their private signals.
Thus for large groups almost all of the private signals are effectively
lost, i.e. not aggregated in the decisions of others. 
\subsubsection{Rational groupthink}

In the proof of this theorem we calculate the asymptotic probability of the
event that \textit{all} agents choose the wrong certainty action in
\textit{almost all time periods} up to time $t$. We call this event
``\textit{rational groupthink}'' and show that its probability is
already high, which implies that the probability that one particular
agent errs at time $t$ is also high. 

When a wrong consensus
forms by chance in the beginning, it is hard to break and can last
for a long time, with surprisingly high probability. To understand why this occurs, we observe that conditioned on a wrong consensus forming, each agent needs a stronger-than-indifferent signal to break the consensus. This is because the private signal needs to to overcome what is learned by observing that the other agents have not broken the consensus. As periods progress, conditioned on the consensus not being broken, the required signal threshold rises and rises. Indeed, after a long time, the threshold will be arbitrarily high. As correct signals are, in the long-run, more likely than incorrect signals, it follows that conditioned on being below the threshold, the agents' signals will be close to it, and in particular will indicate the \textit{correct} action. Thus the private signals of each agent, which initially indicated the wrong action, eventually strongly indicate the correct action, but are still ignored due to the overwhelming information provided by the actions of others. This intuition is formalized in the next result, which is based the large deviation principle that states that when an unlikely even occurs, it is very likely to occur in the most likely way.

%In fact, conditioned on rational groupthink, it holds, with high probability, that the private signals of each agent, which initially indicated the wrong action, eventually strongly indicate the \textit{correct action}, but are still ignored due to the overwhelming information provided by the actions of others. This happens, since, conditioned on a wrong consensus, the threshold of private signal needed to break the consensus rises over time. We thus find the term rational groupthink an apt description of the phenomenon. We formally express this in the following proposition.

Define $\actmin_{t}$ to be the lowest action that is taken by any agent with positive probability at time $t$. In many cases $\actmin_t = \alpha^{\slow}$;
for example, this holds when the private signals are unbounded. For bounded signals this holds for all $t$ large enough, but may not hold for initial $t$.\footnote{Recall that the action $a^i_t$ is chosen according to the posterior belief $p^i_t$. By standard arguments, the set of beliefs in which each possible action is taken is an interval. This induces an order on the actions, from the lowest one, which must be $\alpha^{\slow}$, and is taken for the lowest beliefs, to the highest one, $\alpha^{\shigh}$, which is taken for the highest beliefs.
%Note that when signals are unbounded, then $\actmin_{t}=\alpha^{\slow}$. 
We discuss this technical issue in detail in \S\ref{sec:all_make_mistake}.}%, for example
%if the prior is extreme and the private signals are weak. 

Denote by $\hat p_{i}^{t}=\mathbb{P}\left[\Theta=\shigh\mid s_{1}^{i},\ldots,s_{t}^{i}\right]$
the probability assigned to the good state given only agent $i$'s
signals.
\begin{thm}
\label{prop:groupthink}Condition on the state being good $\Theta=\shigh$. % the state being good. %
In the long run, conditional on all agents taking the eventually \textbf{incorrect}  action $\actmin_t$ in every period, the private signals of every agent strongly
indicate the \textbf{correct} certainty action. That is, for
every  $\varepsilon>0$ it holds that
\[
\lim_{t\to\infty}\mathbb{P}_{\shigh}\left[\hat p_{i}^{t}>1-\varepsilon\text{ for all } i\mid a_{\tau}^{j}=\actmin_\tau\text{ for all }\tau\leq t\text{ and all }j\right]=1.
\]
The analogous statement holds in the bad state.
\end{thm}
Note that Theorem~\ref{prop:groupthink} is \textit{not} a consequence
of the law of large numbers, as conditional on taking the wrong action
the distribution of signals is not independent. Indeed, the result
of Theorem~\ref{prop:groupthink} does not hold in the single
agent case, where\textemdash in sharp contrast\textemdash conditional
on choosing the wrong action the agent holds wrong beliefs. It shows
that in a multi-agent learning problem agents will (with high probability)
have received correct signals even conditioned on choosing the wrong
action. This phenomenon, which does not have an analogue in sequential
herding models, seems striking, as it does not involve irrationality,
and yet results in a group taking an action which contradicts each
and every member's private information.

\subsubsection{Early Period Mistake Probabilities}

Theorem \ref{thm:speed-bound-1} is a statement about asymptotic rates.
In fact, if one were to increase the number of agents while holding
the private signal distributions fixed, the probability of the agents
choosing correctly at any given period $t>1$ approaches 1. Thus,
a more interesting setting is the one studied numerically in \S\ref{sec:leading-example}, and which we analyze formally in this section. In this setting, as we increase the number
of agents, we decrease the informativeness of each agent's signal,
while keeping fixed the amount of information available to all agents
together.

We consider $n$ agents who each receive normal private signals with
fixed conditional means $\pm1$ and variance $n$. If such signals
were publicly observable they would be informationally equivalent
to a single normal signal with variance $1$ each period. In this
setting, Theorem \ref{thm:speed-bound-1} implies that the speed of
learning would be inversely proportional to the size of society, and
in particular would tend to zero as $n$ tends to infinity. 

To test the robustness of this asymptotic speed of learning result,
we perform a detailed analysis of the early periods, showing that,
as the number of agents increases, they learn less and less from each
other's actions. Thus, the asymptotic result of Theorem \ref{thm:speed-bound-1},
which stated that the agents learn little from each other's actions
in the long run, ``kicks in'' early on (in fact, already in the
\textit{second} period), in the sense that with high probability the
agents learn nothing from each other's actions after the first period.
\begin{thm}
\label{thm:short-term}Suppose $n$ agents have a uniform prior, normal private signals
with conditional distributions $\mathcal{N}\left(\pm1,n\right)$ and
want to match the state, %\footnote{See \S \ref{subsec:Matching-the-State}.},
so that $\bar{u}(\theta,a)=\mathbf{1}_{\{a=\theta\}}$. Then, for
every $t$, the probability that all agents in the periods $\left\{ 2,3,\ldots,t\right\} $
choose the action that the majority of the agents chose in period
1 converges to one as $n$ goes to infinity. 
\end{thm}
Note that the theorem statement also holds conditioned on the first period majority taking the wrong action, since this event occurs with probability that is bounded away from zero. Thus the private signals of periods $\left\{ 2,\ldots,t\right\} $
are with high probability not strong enough to induce a deviation
from the first period consensus. Consequently, the actions in these
periods are correct only if the action taken by the majority in the
first period is correct. This probability is bounded by $\Phi(1)\sim0.84$
for any $n$. Of course, this probability can be arbitrarily close
to $1/2$ if the private signal distributions have a larger variance.
The numerical simulations in \S\ref{sec:leading-example} show that a lot of information is lost even for groups of moderate size such as 40 or 100 agents.
%In this case, almost all information is lost even in early periods,
%if the number of agents is sufficiently high. 

The intuition behind this result is the following: after observing
the first round actions, the probability that a particular agent will
have a strong enough signal to deviate from the majority opinion (action)
is small. Increasing the number of agents yields two opposing forces: with more agents and weaker signals for each agent, each particular agent is less likely to deviate from the consensus, but because there are more agents, it is more likely that some agent deviates. It follows from properties of the normal distribution that the probability of receiving a strong signal vanishes more quickly than $\frac{1}{n}$. Since the probability that at least one agent breaks the consensus grows linearly in $n$ (by the union bound), the first effect dominates the second for large $n$. 
%A calculation in the proof of this theorem shows that first effect dominates the second, so that the probability that no agent deviates is almost one. 
When agents
observe that no one has deviated, it further strengthens (if not by
much) their belief in the majority opinion, thus again delaying the
breaking of the consensus. Of course, when the initial consensus is
wrong, eventually it is broken.

%The intuition behind this result is the following: after observing the first round actions, the probability that a particular agent will have a strong enough signal to deviate from the majority opinion (action) is small. In fact, it is so small that the probability that \emph{no }agent deviates is almost one, and moreover it takes many periods until any agent has a strong enough signal to deviate. When agents observe that no one has deviated, it further strengthens (if not by much) their belief in the majority opinion, thus again delaying the breaking of the consensus. Of course, when the initial consensus is wrong, eventually it is broken.

\section{Learning Dynamics\label{sec:Learning-Dynamics}}

In this section we analyze the learning dynamics in detail and explain how we prove the results of \S\ref{sec:result}. We discuss how agents interpret each other's actions and how they choose their own. The analysis of these learning dynamics is related to questions in random walks and requires the application of large deviations techniques. We provide a self-contained introduction to large deviations in the appendix.

\subsection{Preliminaries}

As an agent's expected utility for a given action is linear in her
posterior belief $p_{t}^{i}$, the set of beliefs where she takes
a given action is an interval. It will be convenient to define the
agent's log-likelihood ratio (LLR) 
\begin{equation}
\label{eq:llr-def}
    L_{t}^{i}:=\log \frac{p_{t}^{i}}{1-p_{t}^{i}}\,.
\end{equation}
We define the \textit{private LLR} $R^i_{t}$ as the LLR calculated based only on an agent's private signals. It follows from Bayes' law that
\begin{equation}
    R_{t}^i:=L_{0}^i+\sum_{\tau=1}^{t}\ell_{\tau}^i.\,.\label{eq:LogLikelihood-1}
\end{equation}

As the LLR is a monotone transformation of the agent's posterior belief,
and as a myopic agent's action is determined by her posterior, the
same holds true in terms of LLRs. This can be summarized in the following
lemma. 
\begin{lem}
\label{lem:OptimalActions-1}There exist disjoint intervals $(\underline{L}(\act),\overline{L}(\act))\subset\mathbb{R}\cup\left\{ -\infty,+\infty\right\} $,
one for each action $\act\in A$, such that, with probability one,
$a_{t}^{i}=\act$ if and only if $L_{t}^{i}\in(\underline{L}(\act),\overline{L}(\act))$. 
\end{lem}
Note that we assumed the prior is generic, which rules out indifference.
To characterize the agent's actions it thus suffices to characterize
her LLR. Note, that for the certainty action $\alpha^{\slow}$ it holds
that $\underline{L}(\act^{\slow})=-\infty$, and that analogously $\overline{L}(\act^{\shigh})=+\infty$.

\subsection{Autarky\label{sec:SingleAgent-1}}

As a benchmark, we first describe the classical autarky setting where
a single agent acts by himself. In this section we omit the superscript
signifying the agent.

\subsubsection*{Probability of Mistakes}

As a consequence of Lemma \ref{lem:OptimalActions-1}, the probability
that the agent chooses the wrong action in period $t$ when the state
equals $\theta$ is given by 
\begin{align}
\mathbb{P}_{\theta}\left[a_{t}\neq\act^{\theta}\right] & =\begin{cases}
\mathbb{P}_{\shigh}\left[L_{t}\leq\underline{L}(\act^{\shigh})\right] & \text{ if }\theta=\shigh\\
\mathbb{P}_{\slow}\left[L_{t}\geq\overline{L}(\act^{\slow})\right] & \text{ if }\theta=\slow
\end{cases}\,.\label{eq:ProbMistake}
\end{align}
Hence, to calculate the probability of a mistake one needs to calculate
the probability that the LLR is in a given interval. In the single agent case the private signals are all the available
information, so  $L_{t}=R_{t}$. By (\ref{eq:LogLikelihood-1})
the LLR is the sum of increments which are i.i.d.~conditional on
the state, and hence $\left(L_{t}\right)_{t}$ is a random walk. 

The short-run probability that a random walk is within a given interval is hard to calculate and depends very finely on the distribution of its increments.\footnote{The only exception are a few cases where the distribution of the LLR $L_{t}$ is known in closed form for every $t$, such as the normal case. Even in the normal case it seems to us intractable to calculate in closed form the mistake probability in early periods in the multi-agent case.} As this makes it impossible\textemdash even in the single agent case\textemdash to obtain any general results on the probability that the agent makes a mistake, we focus on the long-run probability of mistakes, which can be analyzed for general signal structures.\footnote{The long-run behavior of random walks has been studied in \emph{large deviations theory}, with one of the earliest results due to \citet{cramer1944large}, who studied these questions in the context of calculating premiums for insurers. We will use some of the ideas and tools from this theory in our analysis; a self-contained introduction is given in Appendix \ref{sec:Appendix-lambda} for the convenience of the reader.}

\subsubsection*{Beliefs}

As $R_t$ is a random walk we can use large deviation theory to estimate the probability that the private
LLR $R_{t}$ deviates from its expectation, conditional on the state.
To this end, recall that $\lambda_{\theta}:\mathbb{R}\to\mathbb{R}$ is the cumulant generating
function of the increments of the LLR in state $\theta$.\footnote{Defined in \S\ref{subsec:results-autarky} by $\lambda_{\shigh}(z)=-\log\,\mathbb{E}_{\shigh}\left[\mathrm{e}^{-z\,\ell}\right]$ and $\lambda_{\slow}(z)=-\log\,\mathbb{E}_{\slow}\left[\mathrm{e}^{z\,\ell}\right]$.} 
Denote its Fenchel conjugate by
\[
\lambda_{\theta}^{\star}(\eta):=\sup_{z\geq0}\lambda_{\theta}(z)-\eta\cdot z.
\]

Given these definition, we are ready to state the basic classical
large deviations estimate that we use in this paper.
\begin{lem}
\label{lem:LLRDeviations}For any $\mathbb{E}_{\slow}\left[\ell\right]<\eta<\mathbb{E}_{\shigh}\left[\ell\right]$
it holds that\footnote{Here each $o(t)$ denotes a different function, so that the first
line can be alternatively written as follows: For every $f_t$
with $\lim_{t\to\infty}f_t/t=0$ there exists a $g_t$ with
$\lim_{t\to\infty}g_t/t=0$ such that $\mathbb{P}_{\shigh}\left[R_{t}\leq\eta\cdot t+f_t\right]=\mathrm{e}^{-\lambda_{\shigh}^{\star}(\eta)\cdot t+g_t}$.} 
\begin{align*}
\mathbb{P}_{\shigh}\left[R_{t}\leq\eta\cdot t+o(t)\right] & =\mathrm{e}^{-\lambda_{\shigh}^{\star}(\eta)\cdot t+o(t)}\\
\mathbb{P}_{\slow}\left[R_{t}\geq\eta\cdot t+o(t)\right] & =\mathrm{e}^{-\lambda_{\slow}^{\star}(-\eta)\cdot t+o(t)}.
\end{align*}
\end{lem}
This Lemma states that the probability that the random walk $R_{t}$
deviates from its (conditional) expectation is exponentially small,
and decays with a rate that can be calculated exactly in terms of
$\lambda_{\shigh}^{\star}$ or $\lambda_{\slow}^{\star}$. The proof of Lemma
\ref{lem:LLRDeviations} in the Appendix uses the properties of $\lambda_{\theta}$
and $\lambda_{\theta}^{\star}$ to verify that the increments of the
LLR process in both states are such that large deviation theory results
are applicable. Lemma \ref{lem:LLRDeviations} allows us to calculate
the probability of a mistake conditional on each state, immediately
implying Fact \ref{thm:ErrorProbSingleAgent-1}, which states that\footnote{We note that it is possible to strengthen this result by replacing
the lower order $o(t)$ term by $O(\log(t))$ using the Bahadur-Rao
exact asymptotics method (see~\citet[Pages 110-113]{dembo1998large}
for a detailed derivation). However, such precision will provide little
additional economic insight while significantly complicating the proofs,
and thus we will not pursue it.} 

\[
\mathbb{P}\left[a_{t}\neq\act^{\Theta}\right]=\mathrm{e}^{-\ra\cdot t+o(t)}\,,
\]
where $\ra=\lambda_{\shigh}^{\star}(0)=\lambda_{\slow}^{\star}(0)$.

\subsection{Many Agents and the Groupthink Effect\label{subsec:GroupThinkExplanation}}

In this section we consider $n\geq2$ agents. Each agent observes
a sequence of private signals $s_{1}^{i},\ldots,s_{t}^{i}$, and the
action taken by the other agents in previous periods $\left(a_{\tau}^{j}\right)_{\tau<t,j\neq i}$.
In this setting we prove Theorem \ref{thm:speed-bound-1}. 

\subsubsection*{The Probability that All Agents Make a Mistake in Every Period}
\label{sec:all_make_mistake}
%To bound the probability of mistake, we consider the event $G_{t}$
%that all agents choose the action $\alpha^{\slow}$ in all time periods
%up to $t$: 
%\[
%G_{t}=\cap_{i=1}^{n}\cap_{\tau=1}^{t}\left\{ a_{\tau}^{i}=\alpha^{\slow}\right\} .
%\]
We define for each $t$ the action $\actmin_{t}$ to be the lowest
action (i.e., having the lowest $\overline{L}(\alpha)$) that is taken
by any agent with positive probability at time $t$, and observe that
$\actmin_{t}$ is equal to $\alpha^{\slow}$ for all $t$ large enough.
To bound the probability of mistake, we consider the event $G_{t}$
that all agents choose the action $\actmin_{t}$ in all time periods
up to $t$:
\[
G_{t}=\{a_{\tau}^{i}=\actmin_\tau\text{ for all }\tau\leq t\text{ and all }i\}.
\]
To simplify the exposition we assume in the main text that $\actmin_{t}=\alpha^{\slow}$.\footnote{This is the case, for example, if the prior is not too extreme relative
to the maximal possible private signal strength, or if the private
signals are unbounded. Otherwise, it may be the case that agents never
take the wrong certainty action in some initial periods, for example
if the prior is extreme and the private signals are weak. In Appendix
\ref{sec:AppendixBidirectional} we drop this assumption and formally show that all our
results also hold in general.} Conditioned on $\Theta=\shigh$, the event $G_{t}$ is the event that
all the agents are, and always have been, in unanimous agreement on
the \emph{wrong} action $\alpha^{\slow}$. We thus call $G_{t}$ the \emph{rational
groupthink} event. The event $G_{t}$ implies that all agents made
a mistake in period $t$, conditioned on $\Theta=\shigh$. Thus calculating
the probability of $G_{t}$ will provide a lower bound on the probability
that a particular agent makes a mistake.

This event can be written as $G_{t}^{1}\cap\cdots\cap G_{t}^{n}$,
where $G_{t}^{i}$ is the event that agent $i$ chooses the wrong
action $\alpha^{\slow}$ in every period $\tau\leq t$. To calculate the
probability of $G_{t}$, it would of course have been convenient if
these $n$ events were independent, conditioned on $\Theta$. However,
due to the fact that the agents' actions are strongly intertwined,
these events are not independent; given that agent 1 played the action
$\alpha^{\slow}$\textemdash which is optimal in the bad state\textemdash in
all previous time periods, agent 2 assigns a higher probability to
the bad state and is more likely to also play the same action. This
poses a difficulty for the analysis of this model, which is a direct
consequence of the fact that the agents' actions are intricately dependent
on their higher order beliefs.

\subsubsection*{Decomposition in Independent Events}

Perhaps surprisingly, it turns out that $G_{t}$ can never-the-less
be written as the intersection of conditionally independent events, one for each agent.
The event associated with agent $i$ is the event that agent $i$'s private LLRs $R^i_1,\ldots,R^i_t$ stay below a time dependent threshold $q_1,\ldots,q_t$ (Lemma~\ref{lem:ThresholdRepresentation}).
This reduces the problem to characterizing the thresholds, which we do after stating this result.

\begin{lem}
\label{lem:ThresholdRepresentation}There exists a sequence of thresholds
$(q_{\tau})_{\tau}$ such that the event $G_{t}$ equals the event
that no agent's private LLR $R^{i}$ hits the threshold $q_\tau$ before
period $t$ 
\[
G_{t}=\bigcap_{i=1}^{n}\{R_{\tau}^{i}\leq q_{\tau}\text{ for all }\tau\leq t\}\,.
\]
\end{lem}
Thus, if we denote
\[
W_{t}^{i}:=\{R_{\tau}^{i}\leq q_{\tau}\text{ for all }\tau\leq t\},
\]
then we have written $G_t = \cap_i W^i_t$ as the intersection of independent events.

The proof of Lemma \ref{lem:ThresholdRepresentation} in Appendix
\ref{sec:AppendixBidirectional} shows this result recursively. Intuitively,
whenever $G_{t-1}$ occurs, all agents took the action $\alpha^{\slow}$
up to time $t-1$. By the induction hypothesis this implies that the
private LLR of all other agents was below the threshold $q_{\tau}$
in all previous periods. As conditional on each state the private
LLR's of different agents are independent, whether agent $i$ takes
the action $\alpha^{\slow}$ at time $t$ conditional on $G_{t-1}$, depends
only on her private LLR $R_{t}^{i}$. As $\alpha^{\slow}$ is the most
extreme action it follows that the set of private LLRs where the agent
takes the action $\alpha^{\slow}$ must be a half-infinite interval and
is thus characterized by a threshold $q_{\tau}$. By symmetry, this
is the same threshold for all agents.

\subsubsection*{Calculating the Thresholds}

We now provide a sketch of the argument 
%(omitting many technical details)
which we use in the appendix to characterize the threshold $q_{t}$.
The threshold $q_{t}$ admits a simple interpretation: it determines
how high a private LLR $R_{t}^{i}$ an agent must have in order to
break from the consensus, and not take action $\alpha^{\slow}$ at time
$t$, after having seen everyone take it so far. To calculate the
$q_{t+1}$'s we consider agent $j$'s decision problem at time $t+1$,
conditioned on $G_{t}$. The information available to her is her own
private signals (summarized in her private log-likelihood ratio $R_{t+1}^{j}$),
and in addition the fact that all other agents have chosen $\alpha^{\slow}$
up to this point. But the latter observation is equivalent to knowing
that all the other agent's private log-likelihood ratios have been
under the thresholds $q_{\tau}$ in all previous time periods. Formally, for agent $j$ to know that $G_{t}$ has occurred, is equivalent to knowing that 
\[
W_{t}^{i}=\{R_{\tau}^{i}\leq q_{\tau}\text{ for all }\tau\leq t\}
\]
has occurred for all agents $i\neq j$.

How does knowing that agent $i$'s private LLR has been below $q_{\tau}$
in all previous periods (i.e.\ $W_{t}^{i}$ occurred) influence agent
$j$'s posterior? To answer this question we consider the log-likelihood
ratio induced by this event, and show that it is asymptotically equal to the logarithm of the probability of the event $R_{t}^{i}\leq q_{t}$,
i.e., the event that agent $i$'s private LLR is below the threshold
$q_{t}$ at just the last period.\footnote{This result is similar in spirit to the
Ballot Theorem of \citet{bertrand1887solution}, which implies that
the probability that a random walk is below a constant threshold in
all prior periods approximately equals (up to sub-exponential terms)
the probability that the random walk is below this threshold in the
last period.}

%
%\begin{equation}
%\log\frac{\mathbb{P}_{\shigh}\left[W_{t}^{i}\right]}%{\mathbb{P}_{\slow}\left[W_{t}^{i}\right]}\,\cdot
%\end{equation}
%
%We show in the appendixthat the rate of the event $W_{t}^{i}$ conditioned on $\Theta=\shigh$ is asymptotically the same as that of the event $R_{t}^{i}\leq q_{t}$, i.e., the event that agent $i$'s private LLR is below the threshold $q_{t}$ at just the last period:\footnote{This result is similar in spirit to the Ballot Theorem of \citet{bertrand1887solution}, which implies that the probability that a random walk is below a constant threshold in all prior periods approximately equals (up to sub-exponential terms) the probability that the random walk is below this threshold in the last period.}
%\[
%\log\mathbb{P}_{\shigh}\left[W_{t}^{i}\right]\approx%\log\mathbb{P}_{\shigh}\left[R_{t}^{i}\leq %q_{t}\right]\,.
%\]
In Lemma~\ref{prop:The-limit-q-exists,} in the appendix we
show that the threshold $q_{t}$ is in fact asymptotically linear, i.e.\ the limit
$\beta=\lim_{t\to\infty}q_t/t$ exists.
%This implies that
%$\log\mathbb{P}_{\shigh}\left[W_{t}^{i}\right]\approx\log\mathbb{P}_{\shigh}\left[R_{t}^{i}\leq q_{t}\right]$.
%Thus, the large deviations estimate given in Lemma \ref{lem:LLRDeviations}
%implies that
We argue that $\mathbb{P}_{\slow}\left[W_{t}^{i}\right]$ is bounded away from zero. Combining this with $\log\mathbb{P}_{\shigh}\left[W_{t}^{i}\right]\approx\log\mathbb{P}_{\shigh}\left[R_{t}^{i}\leq q_{t}\right]$, the linearity of $q$, and the large deviations estimate given in Lemma \ref{lem:LLRDeviations} yields\footnote{Throughout the proof sketch we denote by $\approx$ equality up to terms that are of the order $o(t)$.}
%implies that
\begin{equation}
\log\frac{\mathbb{P}_{\shigh}\left[W_{t}^{i}\right]}{\mathbb{P}_{\slow}\left[W_{t}^{i}\right]}\approx\log\mathbb{P}_{\shigh}\left[W_{t}^{i}\right]\approx\log\mathbb{P}_{\shigh}\left[R_{t}^{i}\leq q_{t}\right]\approx\log\mathbb{P}_{\shigh}\left[R_{t}^{i}\leq \beta\cdot t\right]\approx-\lambda_{\shigh}^{\star}(\beta)\cdot t\,.\label{eq:W-q}
\end{equation}
%As conditional on $\Theta=\slow$, the probability of the event $W_{t}^{i}$ that agent
%$i$ takes the correct action $\alpha^{\slow}$ in every period is strictly
%positive, the LLR induced by the event $W_{t}^{i}$ satisfies
%\[
%\log\frac{\mathbb{P}_{\shigh}\left[W_{t}^{i}\right]}{\mathbb{P}_{\slow}\left[W_{t}^{i}\right]}=\log\mathbb{P}_{\shigh}\left[W_{t}^{i}\right]-\log\mathbb{P}_{\slow}\left[W_{t}^{i}\right]=-\lambda_{\shigh}^{\star}(\beta)\cdot t+o(t).
%\]
%The same equation characterizes also the LLR associated with $W_t^i$ as the probability of the agent taking the action $\alpha^{\slow}$ in every period is strictly positive conditioned on $\Theta=\slow$.
%
Since $G_{t}=\bigcap_{i=1}^{n}W_{t}^{i}$, and since the events $\left(W_{t}^{i}\right)_{i}$ are conditionally independent, we get that when a agent $j$ observes $G_t$, her likelihood ratio will be the sum of $R_t^j$ and $n-1$ times the likelihood ratio of $W_t^i$:
%Since the event $G_{t}$ that the private LLR of every agent is below $q_{\tau}$ in every period prior to $t$ is the intersection of the individual events $G_{t}=\bigcap_{i=1}^{n}W_{t}^{i}$, and since these events $\left(W_{t}^{i}\right)_{i}$ are conditionally independent, we get that the log-likelihood ratio of $G_{t}$ is simply a multiple of the LLR of $W_{t}^{i}$.
%\[
%\log\frac{\mathbb{P}_{\shigh}\left[G_{t}\mid W_t^i\right]}{\mathbb{P}_{\slow}\left[G_{t}\mid W_t^i\right]}=-(n-1)\cdot\lambda_{\shigh}^{\star}(\beta)\cdot t+o(t).
%\]
\begin{equation}
\label{eq:L-R}
    L_{t}^{j}\approx R_{t}^{j}-(n-1)\cdot\lambda_{\shigh}^{\star}(\beta)\cdot t\,.
\end{equation}
%The factor here is $n-1$ rather than $n$, since each agent observes only $n-1$ others. Thus, after observing $G_{t}$, agent $j$'s posterior log-likelihood ratio will be the sum of her private LLR $R_{t}^{j}$ and the LLR induced by observing $G_{t}$. By Lemma \ref{lem:OptimalActions-1}, agent $j$ will therefore take which determines the new threshold $q_{t+1}$.
%
Thus, the threshold for the rational groupthink event at time $t+1$ will satisfy
\begin{equation}
    q_{t+1}\approx\beta \cdot t \approx  (n-1)\cdot\lambda_{\shigh}^{\star}(\beta)\cdot t\,.
\end{equation}
Dividing by $t$ and taking the limit as $t$ tends to infinity yields the following fixed point equation for the slope $\beta$ of the thresholds $(q_\tau)_\tau$ (Lemma~\ref{prop:The-limit-q-exists,}) 
\begin{equation}
\beta=(n-1)\cdot\lambda_{\shigh}^{\star}(\beta).\label{eq:q}
\end{equation}
Note that $\beta$ depends only on the private signal distributions, through
$\lambda_{\shigh}^{\star}.$ Since $\lambda_{\shigh}^{\star}$ is non-negative
and decreasing, this equation will always have a unique solution.
We thus have calculated $\beta$ as the solution of the fixed point
equation (\ref{eq:q}). 
%We have thus calculated $q$: it is the solution of the fixed point
%equation (\ref{eq:q}). 

This fixed point equation has a simple intuiton: if the threshold is too high then it is likely that the
others' private LLRs are below it, and so it is likely that they do
not break the consensus. Thus, an agent gains little information from
observing them agreeing with the consensus, and her threshold for
breaking the consensus will be low. This contradicts the initial assumption
that the threshold is high. Likewise, if the threshold is too low,
then an agent learns a lot by observing the consensus endure, and
thus sets a high threshold for breaking it. The fixed point of (\ref{eq:q})
is the value in which these effects are equal.

Given $\beta$, we can use (\ref{eq:W-q})
to determine the probability of the event $W_{t}^{i}$ that agent
$i$ does not break the consensus. Using the facts that the rational
groupthink event $G_{t}$ satisfies $G_{t}=\bigcap_{i=1}^{n}W_{t}^{i}$
and that the $W_{t}^{i}$'s are conditionally independent, we thus
have that 
\begin{align}
\log \mathbb{P}_{\shigh}\left[G_{t}\right]=\log \left( \mathbb{P}_{\shigh}\left[W_{t}^{i}\right]^{n} \right) = n \log \mathbb{P}_{\shigh}\left[W_{t}^{i}\right] \approx -\beta\cdot\frac{n}{n-1}\cdot t & .\label{eq:g-t-q-1}
\end{align}
Consequently, the rate $\rg$ of the event $G_{t}$ that all agents
take the wrong action in all periods up to time $t$ is 
\begin{align}
    \label{eq:rg}
\rg=\frac{n}{n-1}\beta.
\end{align}
Finally, a convexity argument yields that this rate is bounded by the expected log-likelihood ratio of a single signal: $\rg<\mathbb{E}_{\shigh}\left[\ell\right]$ for
any number of agents (Lemma~\ref{claim:q-bound}). % We provide the proof in the appendix.
As the rational groupthink event implies that all agents make a mistake,
this provides a bound on the speed of learning, conditioned on $\Theta=\shigh$:
\[
\mathbb{P}_{\shigh}\left[a_{t}^{i}\neq\alpha^{\shigh}\right]\geq\mathbb{P}_{\shigh}\left[G_{t}\right]=\mathrm{e}^{-\rg\cdot t+o(t)}.
\]
Performing the corresponding calculation when conditioning on the
bad state, we have proven Theorem~\ref{thm:speed-bound-1}, for $\rmax=\min\left\{ \mathbb{E}_{\shigh}\left[\ell\right],-\mathbb{E}_{\slow}\left[\ell\right]\right\} $.

We note that $\rg$ can often be calculated explicitly. For example,
for normal private signals a straightforward calculation shows that
\[
\rg=4\frac{\left(n-\sqrt{n}\right)^{2}}{\left(n-1\right)^{2}}\ra.
\]
A tedious but straightforward calculation shows that $\rmax=4\ra$.

\section{Incomplete Observation Structures}
\label{sec:incomplete}
So far we have assumed that all agents observe the actions of all others in each period. It is natural to ask how the speed of learning changes when we relax this assumption. We consider two very simple cases, and leave the general case to future work.

\begin{figure}
    \begin{minipage}{.32\textwidth}\centering
        \begin{tikzpicture}[
        roundnode/.style={circle, draw=black!60, fill=black!5, very thick, minimum size=7mm},
        ]
        \node[roundnode]      (player2)                         {Agent 2};
        \node[roundnode]        (player1)       [above=of player2] {Agent 1};
         \node at (0.5,-1,1) {{No Observation}};
    \end{tikzpicture}
    \end{minipage}
    \begin{minipage}{.32\textwidth}\centering
    \begin{tikzpicture}[
        roundnode/.style={circle, draw=black!60, fill=black!5, very thick, minimum size=7mm},
        ]
        \node[roundnode]      (player2)                         {Agent 2};
        \node[roundnode]        (player1)       [above=of player2] {Agent 1};
        \draw[->,line width=3pt] (player1.south) -- (player2.north);
         \node at (0.5,-1,1) {{Unidirectional Observation}};
    \end{tikzpicture}
    \end{minipage}
    \begin{minipage}{.32\textwidth}\centering
    \begin{tikzpicture}[
        roundnode/.style={circle, draw=black!60, fill=black!5, very thick, minimum size=7mm},
        ]
        \node[roundnode]      (player2)                         {Agent 2};
        \node[roundnode]        (player1)       [above=of player2] {Agent 1};
        \draw[<->,line width=3pt] (player1.south) -- (player2.north);
         \node at (0.5,-1,1) {{Bidirectional Observation}};
    \end{tikzpicture}
    \end{minipage}
\caption{Different observability structures we analyze in this section. An arrow from agent $i$ to agent $j$ indicates that agent $i$ can observe agent $j$'s actions.}
\end{figure}

First, we consider just two agents. There are three possible observation structures in this case: when neither observes the other, when both observe each other, and when one observes the other, but not vice versa. We have already treated the first two cases, and here we study the speed of learning in the third case. This speed will now depend on the agent. Of course, the agent who observes nothing but her own private signal will learn as in autarky, and so the new result is the speed of learning of the observing agent. Unsurprisingly, we show that the observing agent learns faster than she would in autarky, as she now has additional information in the form of the actions of the other. The less a priori  obvious result is that the observing agent learns more quickly than she does under the bidirectional observation structure. Thus, in this case, adding another channel of communication between the agents reduces the speed of learning.
\begin{thm}
\label{thm:unidirectional} Consider $2$ agents and two settings of observation structures: either ($\leftrightarrow$) both observe each others' past actions or ($\rightarrow$) agent $1$ observes agent $2$'s past actions, but agent $2$ does not observe agent $1$'s past actions. Denote by $e^\leftrightarrow_t$ the probability that agent $1$'s action $a^1_t$ is not equal to $\alpha^\Theta$ in the first setting, any by $e^\rightarrow_t$ the same probability, in the second setting. Then 
\begin{align*}
 \frac{e^\leftrightarrow_t}{e^\rightarrow_t} \geq \mathrm{e}^{r t + o(t)}
\end{align*}
for some $r > 0$ that depends only on the private signal structure.
\end{thm}
In Theorem \ref{prop:Unidirectional} in the Appendix we compute the exact rate at which agent 1 learns in the unidirectional case. This result might be of independent interest. For example, in the case of normal signals it yields that agent 1 learns as fast as she would learn if she observed
$\frac{9}{16}\approx 56\%$ of agent 2’s private signals, instead of her actions (Corollary~\ref{cor:nor-uni}).

Next, we analyze another simple case: the case of a large group of agents, in which agent $1$ can observe the actions of all others, but no other agent can observe any actions. In this case, we show that the speed of learning of agent $1$ grows linearly with the number of agents she observes. While this result is rather straightforward to understand and prove (as agent $1$ has access to $n-1$ independent sources of information), it highlights the fact that the loss of information in the full observation setting is not due to the fact that agents observe actions rather than signals, but to the interdependence of these actions.
\begin{thm}
\label{prop:observing-actions}Suppose $n$ agents all observe private signals only, except for agent $1$, who additionally observes the others' 
past actions. Given the private signal distributions, there exists a constant $r > 0$, which depends only on the distribution of the private signals, such that for any number of agents 
\[
\mathbb{P}\left[a_{t}^{1}\neq\alpha^{\Theta}\right]\leq\mathrm{e}^{-(n-1)r\cdot t+o(t)}.
\]
\end{thm}

\section{Non-Bayesian Beliefs and Over-Precision Bias}
\label{sec:non-bayesian}
We next relax the assumption that agents form beliefs using Bayes rule. The bias we consider is over-confidence about the precision of an agent's own signals as compared to the other agents' signals.\footnote{This bias seems especially relevant in the context of social learning, where it distorts information aggregation. Its importance has been suggested, for example, by \cite{vives2010information}, exercises 4.7 and 6.7. See \cite{moore2015overprecision} and the references therein.} For tractability we focus on the case of normal signals, and, as in \S\ref{sec:leading-example}, each agent's signal is normally distributed with precision $\nicefrac{1}{n}$ and mean $+1$ or $-1$ depending on the state. While the true precision of each agent's signal is $\nicefrac{1}{n}$, each agent believes that their own signal has precision $c \cdot \nicefrac{1}{n}$ with $c >1$ and all the other agents' signals have precision $\nicefrac{1}{n}$ . We consider the case where agents are sophisticated, that means, are aware of the over-precision bias of others and understand how other agents pick their actions.

\subsection{The Effects of Over-Precision Bias} The over-precision bias of the agents has a direct as well as an indirect effect on the agent's ability to learn the state. The direct effect is straightforward: As agents make a mistake when updating their beliefs they are less likely to chose the correct action. The indirect effect is more subtle: as agents (erroneously) attribute a higher precision to their own signal, they put a higher weight on it when picking their action. As an agent's signals are now more likely to influence his actions, his actions reveal more of his private information. Intuitively, this benefits all other agents and allows them to learn faster. 
\begin{figure}[t]
  \includegraphics[width=\linewidth]{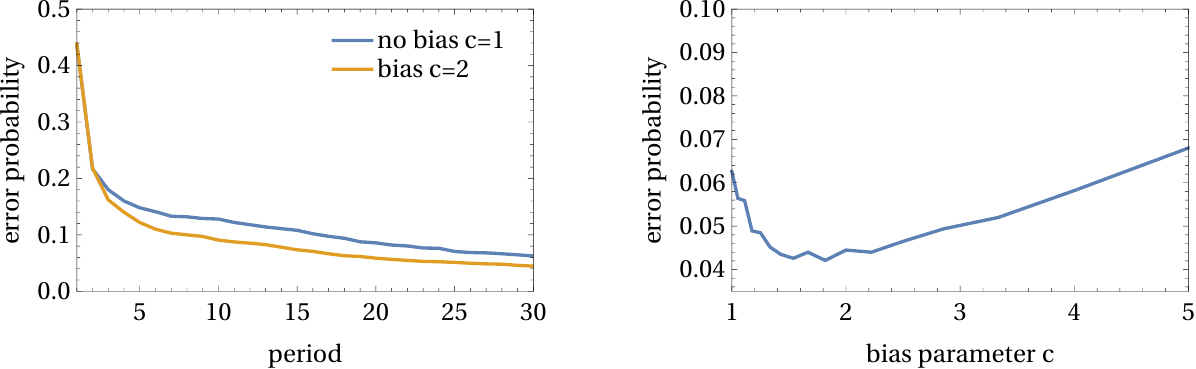}
  \caption{The probability with which an agent takes the wrong action over time in absolute terms for a biased and unbiased agent (on the left) and the error probability in period $t=30$ for various degrees of bias and $n=40$ agents (on the right).}
  \label{fig:biased-learning}
\end{figure}
The right graph of Figure \ref{fig:biased-learning} displays the error probability for various degrees of over-precision in period $30$ in an example with $40$ agents. Maybe surprisingly, agents are less likely to take the wrong action for intermediate biases (in the range between $1$ and $4$). This means that the indirect positive effect coming from the fact that other agents reveal more of their private information dominates the direct effect caused by the deviation from Bayes rule, which leads to wrong beliefs and thus sub-optimal actions. The left graph of Figure \ref{fig:biased-learning} shows how this comparison evolves over time. In the first period, the agent observed only his own signal and the over-precision bias thus has no effect. In the second period there is no positive effect of the over-precision bias as other agents reveal exactly the same information in the first period independent of the bias. The error probability is thus higher---but only slightly---with the bias ($21.7\%$ vs $21.8\%$ for $c=1$ vs $c=2$). However, already from the third period the indirect benefit is larger than the direct loss, leading to lower error probabilities for the biased agents ($18\%$ vs $16\%$ in period 3, and, for example, $13\%$ vs $9\%$ in period 10). 

We conjecture that  for an appropriate choice of the bias parameter $c$, the asymptotic error probability is smaller for biased agents than it is for rational agents, as suggested by Figure~\ref{fig:biased-learning}. Indeed, a straightforward modification of the proof of Theorem~\ref{thm:speed-bound-1} shows that the rate of the groupthink event indeed can increase (implying lower probability of the groupthink event) for biased agents. However, the probability of groupthink provides only a lower bound on the error probability, which we do not know how to explicitly calculate. We thus leave this conjecture for future work.

%
% Social Planner
%
\subsection{A Social Planners Perspective}

An interesting question is which strategies a social planner would pick for the agents in order to maximize the long-run probability with which they pick the right action. The main trade-off faced by such a social planner is that taking a sub-optimal action today increases the mistake probability today, but potentially leads the agent to reveal more information which benefits other agents in the future. While in equilibrium agents do not take this positive informational externality into account, a forward-looking social planner would, and thus potentially has an incentive to intervene with the agents' actions. While solving for the optimal policy is beyond the scope of this paper, the numerical simulations of the previous section already provide some insight into this question: The simulations indicate that agents learn faster when they over-weigh their own signals, which suggests that a social planner could improve welfare by instructing agents to use the \textsl{non-Bayesian} biased updating rule described in the previous section. Thus, while biased learning is suboptimal for myopic individuals, it might be socially beneficial. 

%
% Conclusion
%
\section{Conclusion}

We show that rational groupthink occurs in a complex
environment of agents who observe each other and take actions repeatedly.
As a result, almost all information is lost when the group of agents
is large. We use asymptotic rates as a measure of the speed of learning.
As a robustness test, we show that the same effect holds also in the
early periods, for the case of normal signals.

This article leaves many open questions which could potentially be
analyzed using our approach. 
\begin{enumerate}
    \item We think that it may be feasible to extend our methods beyond the two state case to an arbitrary, finite number of states. This will require the use of high-dimensional large deviation techniques, as the beliefs are now multi-dimensional.
    \item The extension to the case in which different agents have different signals seems more straightforward. We conjecture that the methodology we developed could be used in this setting and will lead to similar conclusions.
    
    \item What happens when the state changes over time? This setting is potentially very interesting, as one could derive steady-state results instead of asymptotic results. We conjecture that results that are similar in spirit will hold, with large groups not performing significantly better than single agents. A major challenge in the analysis is that, since the probability of taking the wrong action does not vanish over time, large deviation techniques no longer apply. Social learning with a changing state and short-lived agents has been studied by \cite{moscarini1998social}, \cite{frongillo2011social} and recently \cite{dasaratha2018social}.
    
    \item What happens with payoff externalities, for example when agents have incentive to coordinate?
    \item What is the optimal policy of a forward-looking social planner who cannot transfer information between the agents? It is unclear to us how one could approach this problem.
    \item Of particular interest is the study of more complex societal structures: how fast do agents learn for a given arbitrary network of observation, which is not the complete network? We briefly tackle some particularly simple examples in \S\ref{sec:incomplete}, but our techniques break down in the general case, as they rely on the fact that the groupthink event is common knowledge.
\end{enumerate}

\bibliographystyle{plainnat}
\bibliography{references}
\newpage{}

\appendix
%dummy comment inserted by tex2lyx to ensure that this paragraph is not empty%dummy comment inserted by tex2lyx to ensure that this paragraph is not empty

\section{The cumulant Generating Functions, their Fenchel Conjugates, and
Large Deviations Estimates\label{sec:Appendix-lambda}}

The long-run behavior of random walks has been studied in large deviations
theory. In this section we first introduce some well known tools from this literature, which will be crucial to understanding the long-run behavior of agents. In the end of this section we derive a sample path large deviation theorem which will be the main tool in our analysis. The proof of this theorem follows well known techniques \citep[see][Chapter 5]{dembo1998large}.

\subsection*{Large Deviations of Random Walks}

Let $X_{1},X_{2},\ldots$ be i.i.d random variables with $\mathbb{E}\left[X_{t}\right]=\mu$
and $Y_{t}=\sum_{\tau=1}^{t}X_{t}$ the associated random walk with
steps $X_{t}.$ By the law of large numbers we know that $Y_{t}$
should approximately equal $\mu\cdot t$. Large deviation theory characterizes
the probability that $Y_{t}$ is much lower, and in particular smaller
than $\eta\cdot t$, for some $\eta<\mu$. Under some technical conditions,
this probability is exponentially small, with a rate $\lambda^{\star}(\eta)$:
\[
\mathbb{P}\left[Y_{t}<\eta\cdot t+o(t)\right]=\mathrm{e}^{-\lambda^{*}(\eta)\cdot t+o(t)}\,,
\]
or equivalently stated 
\[
\lim_{t\to\infty}-\frac{1}{t}\log\mathbb{P}\left[Y_{t}<\eta\cdot t+o(t)\right]=\lambda^{\star}(\eta).
\]
The rate $\lambda^{\star}$ can be calculated explicitly and is the
\emph{Fenchel Conjugate} of the \emph{cumulant generating function}
of the increments

\[
\lambda^{\star}(\eta):=\sup_{z\geq0}\left(-\log\,\mathbb{E}\left[\mathrm{e}^{-z\,X_{1}}\right]-\eta\cdot z\right).
\]
The first proof of a ``large deviation'' result of this flavor is
due to \citet{cramer1944large}, who studied these questions in the
context of calculating premiums for insurers. A standard textbook
on large deviations theory is \citet{dembo1998large}.

In this section we provide an independent proof of this classical
large deviations result, and prove a more specialized one suited to
our needs. We consider a very general setting: we make no assumptions
on the distribution of each step $X_{t}$, and in particular do not
need to assume that it has an expectation.

Denoting $X=X_{1}$, The cumulant generating function $\lambda$ is
(up to sign, as compared to the usual definition) given by

\[
\lambda(z)=-\log\,\mathbb{E}\left[\mathrm{e}^{-z\,X}\right].
\]
Note that when the right hand side is not finite it can only equal
$-\infty$ (and never $+\infty$). 
\begin{lem}
\label{prop:Prop2}$\lambda$ is finite on an interval $I,$ on which
it is concave and on whose interior it is smooth (that is, having
continuous derivatives of all orders). 
\end{lem}
\begin{proof}
Note that $I$ contains $0$, since $\lambda(0)=0$ by definition.
Assume $\lambda(a)$ and $\lambda(b)$ are both finite. Then for any
$r\in(0,1)$ 
\[
\lambda(r\cdot a+(1-r)\cdot b)=-\log\,\mathbb{E}\left[\mathrm{e}^{-(r\cdot a+(1-r)\cdot b)\cdot X}\right]=-\log\,\mathbb{E}\left[\left(\mathrm{e}^{-a\cdot X}\right)^{r}\cdot\left(\mathrm{e}^{-b\cdot X}\right)^{1-r}\right],
\]
which by Hölder's inequality is at least $r\cdot\lambda(a)+(1-r)\cdot\lambda(b)$.
Hence $\lambda$ is finite and concave on a convex subset of $\mathbb{R}$,
or an interval. We omit here the technical proof of smoothness; it
can be found, for example, in \citet[Theorem 1.4.16]{stroock2013mathematics}. 
\end{proof}
It also follows that unless the distribution of $X$ is a point mass
(which is a trivial case), $\lambda$ is strictly concave on $I$.
We assume this henceforth. Note that it could be that $I$ is simply
the singleton $[0,0]$. This is not an interesting case, and we will
show later that in our setting $I$ is larger than that.

The Fenchel conjugate of $\lambda$ is given by 
\[
\lambda^{\star}(\eta)=\sup_{z\geq0}\lambda(z)-\eta\cdot z.
\]
We note a few properties of $\lambda^{\star}$. First, since $\lambda(0)=0$
and $\lambda(z)<\infty$, $\lambda^{\star}$ is well defined and non-negative
(but perhaps equal to infinity for some $\eta$). Second, since $\lambda$
is equal to $-\infty$ whenever it is not finite, the supremum is
attained on $I$, unless it is infinity. Third, since $\lambda$ is
strictly concave on $I$, $\lambda(z)-\eta\cdot z$ is also strictly concave
there, and so the supremum is a maximum and is attained at a single
point $z^*\in I$ whenever it is finite. Additionally, since $\lambda$
is smooth on $I$, this single point $z^*$ satisfies $\lambda'(z^*)=\eta$
if $z^*>0$ (equivalently, if $\lambda^{\star}(\eta)>0)$. I.e., if
$\lambda'(z^*)=\eta$ for some $z^*$ in the interior of $I$ then 
\begin{equation}
\lambda^{*}(\eta)=\lambda(z^*)-\eta\cdot z^*.\label{eq:lambda-star-eta}
\end{equation}
Finally, it is immediate from the definition that $\lambda^{*}$ is
weakly decreasing, and it is likewise easy to see that it is continuous.
This, together with (\ref{eq:lambda-star-eta}) and the fact that
$\lambda'$ is decreasing, yields that $\lambda^{\star}(\eta)=\lambda(0)=0$
whenever $\eta\geq\sup_{z\geq0}\lambda'(z)$. We summarize this in
the following lemma. 
\begin{lem}
\label{prop:lambda-star}Let $I$ be the interval on which $\lambda$
is finite, and let $I^{\star}=\left\{ \eta\,:\,\exists z\in\mathrm{int}I\mbox{ s.t. }\lambda'(z)=\eta\right\} $.
Then 
\begin{enumerate}
\item $\lambda^{*}$ is continuous, non-negative and weakly decreasing.
It is positive and strictly decreasing on $I^{*}$. 
\item $\lambda^{\star}(\eta)=0$ whenever $\eta\geq\sup_{z\geq0}\lambda'(z).$ 
\item If $\eta\in I^{\star}$ and $\lambda'(z^*)=\eta$ then $\lambda^{*}(\eta)=\lambda(z^*)-\eta\cdot z^*$. 
\end{enumerate}
\end{lem}
Given all this, we are ready to state and prove our first large deviations
theorem. 
\begin{thm}[\citealp{cramer1944large}]
\label{thm:large-dev-1}For every $\eta$ such that $\eta>\inf_{z\in I}\lambda'(z)$
it holds that

\[
\mathbb{P}\left[Y_{t}\leq\eta\cdot t+o(t)\right]=\mathrm{e}^{-\lambda^{\star}(\eta)\cdot t+o(t)}.
\]
\end{thm}
\begin{proof}
For the upper bound, we use a Chernoff bound strategy: for any $z\geq0$

\[
\mathbb{P}\left[Y_{t}\leq\eta\cdot t+o(t)\right]=\mathbb{P}\left[\mathrm{e}^{-z\,Y_{t}}\geq\mathrm{e}^{-z\cdot(\eta\cdot t+o(t))}\right],
\]
and so by Markov's inequality

\[
\mathbb{P}\left[Y_{t}\leq\eta\cdot t+o(t)\right]\leq\frac{\mathbb{E}\left[\mathrm{e}^{-z\,Y_{t}}\right]}{\mathrm{e}^{-z\cdot(\eta\cdot t+o(t))}}\cdot
\]
Now, note that $\mathbb{E}\left[\mathrm{e}^{-z\,Y_{t}}\right]=\mathrm{e}^{-\lambda(z)\cdot t}$,
and so 
\[
\mathbb{P}\left[Y_{t}\leq\eta\cdot t+o(t)\right]\leq\mathrm{e}^{-(\lambda(z)-z\cdot\eta)\cdot t+z\cdot o(t)}.
\]
Choosing $z\geq0$ to maximize the coefficient of $t$ yields

\[
\mathbb{P}\left[Y_{t}\leq\eta\cdot t+o(t)\right]\leq\mathrm{e}^{-\lambda^{\star}(\eta)\cdot t+o(t)},
\]
which is the desired upper bound.

We now turn to proving the lower bound. Denote by $\nu$ the law of
$X$, and for some fixed $z$ in the interior of $I$ (to be determined
later) define the probability measure $\tilde{\nu}$ by

\[
\frac{\mathrm{d\tilde{\nu}}}{\mathrm{d}\nu}(x)=\frac{\mathrm{e}^{-zx}}{\mathbb{E}\left[\mathrm{e}^{-zX}\right]}=\mathrm{e}^{\lambda(z)-zx},
\]
and let $\tilde{X}_{t}$ be i.i.d. random variables with law $\tilde{\nu}$.
Note that 
\[
\mathbb{E}\left[\tilde{X}\right]=\frac{\mathbb{E}\left[X\mathrm{e}^{-zX}\right]}{\mathbb{E}\left[\mathrm{e}^{-zX}\right]}=\lambda'(z).
\]
Now, fix any $\eta_{1},\eta_{2}$ such that $\eta_{1}<\eta_{2}<\eta$
and $\lambda'(z)=\eta_{2}$ for some $z$ in the interior of $I$;
this is possible since $\eta>\inf_{z\in I}\lambda'(z)$. This is the
$z$ we choose to take in the definition of $\tilde{\nu}.$ If we
think of $\eta_{2}$ as being close to $\eta$ then the expectation
of $\tilde{X}$, which is equal to $\eta_{2}$, is close to $\eta$.
We have thus ``tilted'' the random variable $X$, which had expectation
$\mu$, to a new random variable with expectation close to $\eta$.

We can bound 
\[
\mathbb{P}\left[Y_{t}\leq\eta\cdot t+o(t)\right]\geq\mathbb{P}\left[\eta_{1}\cdot t\leq Y_{t}\leq\eta\cdot t+o(t)\right]=\int_{\eta_{1}t}^{\eta t+o(t)}1\,\mathrm{d}\nu^{(t)},
\]
where $\nu^{(t)}$ is the $t$-fold convolution of $\nu$ with itself,
and hence the law of $Y_{t}$. It is easy to verify\footnote{See, e.g., \citet[Page 74]{durrett1996probability} or note that the
Radon-Nikodym derivative between the law of $X$ and $\tilde{X}$
is $\mathrm{e}^{zx-\lambda(z)}$, and so the derivative between the
laws of $(X_{1},\ldots,X_{t})$ and $(\tilde{X}_{1},\ldots,\tilde{X}_{t})$
is $\mathrm{e}^{z(x_{1}+\cdots+x_{t})-\lambda(z)\cdot t}$.} that $\mathrm{d}\nu^{(t)}(y)=\mathrm{e}^{zy-\lambda(z)\cdot t}\,\mathrm{d}\tilde{\nu}^{(t)}(y)$,
and so 
\[
\int_{\eta_{1}t}^{\eta t+o(t)}1\,\mathrm{d}\nu^{(t)}=\mathrm{e}^{-\lambda(z)\cdot t}\int_{\eta_{1}t}^{\eta t+o(t)}\mathrm{e}^{zy}\,\mathrm{d}\tilde{\nu}^{(t)}(y),
\]
which we can bound by taking the integrand out of the integral and
replacing $y$ with the lower integration limit: 
\[
\mathrm{e}^{-\lambda(z)\cdot t}\int_{\eta_{1}t}^{\eta t+o(t)}\mathrm{e}^{zy}\,\mathrm{d}\tilde{\nu}^{(t)}(y)\geq\mathrm{e}^{(\eta_{1}z-\lambda(z))\cdot t}\int_{\eta_{1}t}^{\eta t+o(t)}\mathrm{1}\,\mathrm{d}\tilde{\nu}^{(t)}.
\]
Since the law of $\tilde{Y}_{t}=\sum_{\tau=1}^{t}\tilde{X}_{t}$ is
$\tilde{\nu}^{(t)}$, this is implies that
\[
\mathrm{e}^{(\eta_{1}z-\lambda(z))\cdot t}\int_{\eta_{1}t}^{\eta t+o(t)}\mathrm{1}\,\mathrm{d}\tilde{\nu}^{(t)}=\mathrm{e}^{(\eta_{1}z-\lambda(z))\cdot t}\mathbb{P}\left[\eta_{1}\cdot t\leq\tilde{Y}_{t}\leq\eta\cdot t+o(t)\right].
\]
Since $\eta_{1}<\mathbb{E}[\tilde{X}]<\eta$ we have that
$\lim_{t\to \infty}\mathbb{P}\left[\eta_{1}\cdot t\leq\tilde{Y}_{t}\leq\eta\cdot t+o(t)\right]=1$,
by the law of large numbers. Hence 
\begin{align}
    \label{eq:Y_t}
    \liminf_{t\to\infty}\frac{1}{t}\log\mathbb{P}\left[Y_{t}\leq\eta\cdot t+o(t)\right]\geq\eta_{1}z-\lambda(z)
\end{align}
which, by (\ref{eq:lambda-star-eta}), and recalling that $z=(\lambda')^{-1}(\eta_{2})$,
can be written as 
\[
\liminf_{t\to\infty}\frac{1}{t}\log\mathbb{P}\left[Y_{t}\leq\eta\cdot t+o(t)\right]\geq-\lambda^{*}(\eta_{2})-\left(\eta_{2}-\eta_{1}\right)\cdot(\lambda')^{-1}(\eta_{2}).
\]
Taking the limit as $\eta_{1}$ approaches $\eta_{2}$ yields 
\begin{equation}
\liminf_{t\to\infty}\frac{1}{t}\log\mathbb{P}\left[Y_{t}\leq\eta\cdot t+o(t)\right]\geq-\lambda^{*}(\eta_{2}).\label{eq:eta-under}
\end{equation}
We now consider two cases. First, assume that $\eta\leq\sup_{z\geq0}\lambda'(z)$.
In this case we can choose $\eta_{2}$ arbitrarily close to $\eta$,
and by the continuity of $\lambda^{*}$ we get that
\[
\liminf_{t\to\infty}\frac{1}{t}\log\mathbb{P}\left[Y_{t}\leq\eta\cdot t+o(t)\right]\geq-\lambda^{\star}(\eta),
\]
or equivalently 
\[
\mathbb{P}\left[Y_{t}\leq\eta\cdot t+o(t)\right]\geq\mathrm{e}^{-\lambda^{\star}(\eta)\cdot t+o(t)}.
\]

The second case is that $\eta>\sup_{z\geq0}\lambda'(z)$. In this
case $\lambda^{\star}(\eta)=0$ (Lemma~\ref{prop:lambda-star}).
Also, (\ref{eq:eta-under}) holds for any $\eta_{2}<\sup_{z}\lambda'(z)$
and thus it holds for $\eta_{2}=\sup_{z\geq0}\lambda'(z)$. But then
$\lambda^{\star}(\eta_{2})=0=\lambda^{\star}(\eta)$, and so we again
arrive at the same conclusion. 
\end{proof}

\subsection{Sample Path Large Deviation Bounds}
\label{sec:Appendix-Large-Deviation-Pathwise}
In this section we prove a large deviation result that is similar in spirit, and in some sense is stronger
than Theorem~\ref{thm:large-dev-1}, as it shows that the same rate applies to the event
that the sum is below the threshold at all time periods prior to $t$,
rather than just at period $t$. It furthermore does not require the
threshold to be linear, but only asymptotically and from one direction;
both of these generalizations are important. This theorem is similar in spirit to other sample path large deviation results \citep[see, e.g.,][Chapter 5]{dembo1998large}.
\begin{thm}
\label{thm:large-deviation-2}For every $\eta$ such that $\eta>\inf_{z\in I}\lambda'(z)$,
and every sequence $\left( y_{t}\right) _{t\in\mathbb{N}}$ with
$\liminf_{t\to \infty}y_{t}/t=\eta$ and $\mathbb{P}\left[Y_{t}\leq y_{t}\right]>0$
it holds that

\[
\mathbb{P}\left[\cap_{\tau=1}^{t}\left\{ Y_{\tau}\leq y_{\tau}\right\} \right]=\mathrm{e}^{-\lambda^{\star}(\eta)\cdot t+o(t)}.
\]
\end{thm}
\begin{proof}
Let $E_{t}$ be the event $\cap_{\tau=1}^{t}\left\{ Y_{\tau}\leq y_{\tau}\right\} $.
Let $\left( t_{k}\right)$ be a sequence such that $\lim_{k\to \infty}y_{t_{k}}/t_{k}=\eta.$
For every $t$ let $t'$ be the largest $t_{k}$ with $t_{k}\leq t$.
Then by inclusion we have that 
\[
\frac{1}{t}\log\mathbb{P}\left[E_{t}\right]\leq\frac{1}{t'}\log\mathbb{P}\left[Y_{t'}\leq y_{t'}\right].
\]
Using the same Chernoff bound strategy of the proof of Theorem \ref{thm:large-dev-1},
we get that 
\[
\frac{1}{t}\log\mathbb{P}\left[E_{t}\right]\leq-\lambda^{\star}\left(y_{t'}/t'\right).
\]
The continuity of $\lambda$ implies that taking the limit superior
of both sides yields 
\[
\limsup_{t \to \infty}\frac{1}{t}\log\mathbb{P}\left[E_{t}\right]\leq-\lambda^{\star}\left(\eta\right),
\]
or 
\[
\mathbb{P}\left[E_{t}\right]\leq\mathrm{e}^{-\lambda^{\star}(\eta)\cdot t+o(t)}.
\]
To show the other direction, define (as in the proof of Theorem \ref{thm:large-dev-1})
$\tilde{X}_{t}$ to be be i.i.d. random variables with law $\tilde{\nu}$
given by 
\[
\frac{\mathrm{d}\tilde{\nu}}{\mathrm{d}\nu}(x)=\mathrm{e}^{\lambda(z)-zx},
\]
where $\nu$ is the law of $X$, and $z\in I$ is chosen so that $\lambda'(z)=\eta_{2}$
for some $\eta_{1}<\eta_{2}<\eta$, so that the expectation of $\tilde X_t$ is $\eta_2$. It follows from inclusion that 
\[
\mathbb{P}\left[E_{t}\right]\geq\mathbb{P}\left[E_{t}\cap\left\{ Y_{t}\geq \eta_1 \cdot t\right\} \right].
\]
Now, the Radon-Nikodym derivative between the laws of $(X_{1},\ldots,X_{t})$
and $(\tilde{X}_{1},\ldots,\tilde{X}_{t})$ is $\mathrm{e}^{z(x_{1}+\cdots+x_{t})-\lambda(z)\cdot t}$.
Hence 
\[
\mathbb{P}\left[E_{t}\right]\geq\mathbb{E}\left[1_{E_{t}}\cdot1_{Y_{t}\geq \eta_1 \cdot t}\right]=\mathbb{E}\left[1_{\tilde{E}_{t}}\cdot1_{\tilde{Y}_{t}\geq \eta_1 \cdot t}\cdot\mathrm{e}^{z\tilde{Y}_{t}-\lambda(z)\cdot t}\right],
\]
where $\tilde{E}_{t}$ is the event $\cap_{\tau=1}^{t}\{ \tilde{Y}_{\tau}\leq y_{\tau}\} $.
We can bound this expression by taking $\mathrm{e}^{z\tilde{Y}_{t}-\lambda(z)\cdot t}$
out of the integral and replacing it with the lower bound $\eta_1\cdot t$.
This yields 
\begin{align}
\label{eq:E_t}
\mathbb{P}\left[E_{t}\right]\geq\mathrm{e}^{z(\eta_1 \cdot t)-\lambda(z)\cdot t}\cdot\mathbb{P}\left[\tilde{E}_{t}\cap\left\{ \tilde{Y}_{t}\geq \eta_1 \cdot t\right\} \right].
\end{align}
Since the expectation of $\tilde{Y_{t}}/t$ is strictly higher than $\eta_1$, we have that $\lim_{t \to \infty}\mathbb{P}\big[\tilde{Y}_{t}\geq \eta_1 \cdot t\big]=1$
by the weak law of large numbers. We claim that $\lim_{t \to \infty}\mathbb{P}[\tilde E_t]>0$, and show this below. Thus $\lim_{t}\mathbb{P}[\tilde{E}_{t}\cap\{ \tilde{Y}_{t}\geq \eta_1 \cdot t\} ]>0$. 
We can therefore deduce from \eqref{eq:E_t} that
\begin{align*}
\liminf_{t \to \infty}\frac{1}{t}\log\mathbb{P}\left[E_{t}\right]
&\geq z\cdot\eta_{1}-\lambda(z) + \lim_{t\to\infty} \frac{1}{t} \log  \mathbb{P}\left[\tilde{E}_{t}\cap\left\{ \tilde{Y}_{t}\geq \eta_1 \cdot t\right\} \right]\\
&\geq z\cdot\eta_{1}-\lambda(z)\,.
\end{align*}
Proceeding as in the proof of Theorem \ref{thm:large-dev-1} following equation \eqref{eq:Y_t} yields
that 
\[
\mathbb{P}\left[E_{t}\right]\geq\mathrm{e}^{-\lambda^{\star}(\eta)\cdot t+o(t)}\,,
\]
which is what we set out to prove.

It thus remains to be shown that $\lim_{t\to \infty}\mathbb{P}[\tilde{E}_{t}]>0$. Recall that $\tilde{E}_{t}=\cap_{\tau=1}^{t}\{ \tilde{Y}_{\tau}\leq y_{\tau}\} $ is the event that $\tilde Y_t$ is under the threshold $y_t$ up to time $t$. Since $\tilde E_{t+1} \subseteq \tilde E_t$, we need to show that the probability of $\tilde E := \cap_{t=1}^\infty \tilde E_t$ is positive. This is the event that $\tilde Y_t$ is always under the threshold $y_t$.

Denote $\tilde F_t = \cap_{\tau={t+1}}^\infty \{ \tilde{Y}_{\tau}\leq y_{\tau}\}$. This is the event that $Y_t$ is under the threshold $y_t$ after time $t$. Thus, for any $t$,  $\tilde E = \tilde E_{t} \cap \tilde F_{t}$. Recall also that $\liminf_t y_t/t=\eta$, and that $\mathbb{E}[\tilde Y_t/t] = \eta_1 < \eta$. It thus follows from the strong law of large numbers that $Y_t \leq y_t$  eventually: $\lim_t\mathbb{P}[\tilde F_t]=1$. In particular there is some $t_0$ such that $\mathbb{P}[\tilde F_{t_0}]>0$.

An hypothesis of this theorem is that $\mathbb{P}[\tilde Y_t \leq y_t]>0$ for all $t$. As these events are all positively correlated, it is easy to show that the intersection of any finite number of them has positive probability, and in particular that $\mathbb{P}[\tilde{E}_{t_0}]>0$. By the same reasoning $\tilde{E}_{t_0}$ is positively correlated with $\tilde F_{t_0}$, and therefore
$$
\lim_{t\to \infty}\mathbb{P}[\tilde{E}_{t}]=
\mathbb{P}[\tilde E]=\mathbb{P}[\tilde E_{t_0}\cap \tilde F_{t_0}] \geq \mathbb{P}[\tilde E_{t_0}] \cdot \mathbb{P}[\tilde F_{t_0}] > 0.
$$
\end{proof}

\section{Application of Large Deviation Estimates}

\label{app:application-of-large-dev}In this section we prove a number
of claims regarding the functions $\lambda_{\theta}$ and $\lambda_{\theta}^{*}$.
Recall that for $\theta\in\left\{ h,l\right\} $

\[
\lambda_{\shigh}(z):=-\log\,\mathbb{E}_{\shigh}\left[\mathrm{e}^{-z\,\ell}\right]\quad\quad\lambda_{\slow}(z):=-\log\,\mathbb{E}_{\slow}\left[\mathrm{e}^{z\,\ell}\right]\,,
\]
where $\ell$ is a random variable with the same law as any $\ell_{t}^{i}$,
and 
\[
\lambda_{\theta}^{\star}(\eta)=\max_{z}\lambda_{\theta}(z)-\eta\cdot z.
\]

We first note that by the definition of $\lambda_{\theta}$ we have
that

\begin{equation}
\lambda_{\shigh}(z)=-\log\int\exp\left(-z\cdot\log\frac{\mathrm{d\mu_{\shigh}}}{\mathrm{d}\mu_{\slow}}(s)\right)\mathrm{d}\mu_{\shigh}(s)=-\log\int\left(\frac{\mathrm{d\mu_{\slow}}}{\mathrm{d}\mu_{\shigh}}(s)\right)^{z}\mathrm{d}\mu_{\shigh}(s).\label{eq:lambda-h}
\end{equation}
It follows immediately that there is a simple connection between $\lambda_{\shigh}$
and $\lambda_{\slow}$

\[
\lambda_{\slow}(z)=\lambda_{\shigh}(1-z).
\]
Furthermore, as for every $\eta$ between $\mathbb{E}_{\shigh}\left[\ell\right]$
and $\mathbb{E}_{\slow}\left[\ell\right]$ the maximum in the definition
of $\lambda_{\shigh}^{\star}$ is achieved for some $z\in(0,1)$, it follows
that there is also a simple connection between $\lambda_{\shigh}^{\star}$
and $\lambda_{\slow}^{\star}:$
\begin{equation}
\lambda_{\slow}^{\star}(\eta)=\lambda_{\shigh}^{\star}(-\eta)-\eta.\label{eq:lambda-h-lambda-l}
\end{equation}
We will accordingly state some results in terms of $\lambda_{\shigh}$
and $\lambda_{\shigh}^{\star}$ only. It also follows from (\ref{eq:lambda-h})
that the interval $I$ on which $\lambda_{\shigh}$ is finite contains
$[0,1]$. Since from the definitions we have that $\lambda_{\shigh}'(0)=\mathbb{E}_{\shigh}\left[\ell\right]$,
and since $\lambda_{\shigh}'(1)=\mathbb{E}_{\slow}\left[\ell\right]$ by the
relation between $\lambda_{\shigh}$ and $\lambda_{\slow}$, we have shown
the following lemma.
\begin{lem}
\label{lem:lambda-and-lambda-star}$\lambda_{\theta}(z)$ and $\lambda_{\theta}^{\star}(\eta)$
are finite for all $z\in[0,1]$ and $\eta\in(\mathbb{E}_{\slow}\left[\ell\right],\mathbb{E}_{\shigh}\left[\ell\right])$.
Furthermore, 
\begin{equation}
\lambda_{\shigh}(z)=\lambda_{\slow}(1-z)\text{ and }\lambda_{\shigh}^{*}(\eta)=\lambda_{\slow}^{*}(-\eta)-\eta\,.\label{eq:LegendreSym-1}
\end{equation}
\end{lem}
\begin{proof}[Proof of Lemma \ref{lem:LLRDeviations}]
Given Lemma \ref{lem:lambda-and-lambda-star}, Lemma \ref{lem:LLRDeviations}
is an immediate corollary of Theorem \ref{thm:large-dev-1}.
\end{proof}
The following simple observation will be useful on several occasions: 
\begin{lem}
\label{claim:r^a-lambda}Let $\ra=\lambda_{\shigh}^{\star}(0).$ Then $\ra=\max_{z\in(0,1)}\lambda_{\shigh}(z)=\max_{z\in(0,1)}\lambda_{\slow}(z)=\lambda_{\slow}^{*}(0)$,
$\ra<\min\left\{ \mathbb{E}_{\shigh}\left[\ell\right],-\mathbb{E}_{\slow}\left[\ell\right]\right\} $,
and $\min\left\{ \lambda_{\shigh}^{\star}(\ra),\lambda_{\slow}^{\star}(\ra)\right\} >0.$ 
\end{lem}
\begin{proof}
That $\ra=\max_{z\in(0,1)}\lambda_{\shigh}(z)=\max_{z\in(0,1)}\lambda_{\slow}(z)=\lambda_{\slow}^{*}(0)$
follows immediately from the definitions. Now, note that $\mathbb{E}_{\shigh}\left[\ell\right]=\lambda_{\shigh}'(0)$.
Thus $\ra<\mathbb{E}_{\shigh}\text{\ensuremath{\left[\ell\right]}}$ is
a simple consequence of the fact that $\ra=\lambda_{\shigh}^{\star}(0)=\max_{z\geq0}\lambda(z)$,
that this maximum is obtained in $(0,1)$, and that $\lambda_{\shigh}$
is strictly concave. It follows from the same considerations that
$\ra<-\mathbb{E}_{\slow}\left[\ell\right].$ Finally, by Lemma~\ref{prop:lambda-star},
$\lambda_{\shigh}^{\star}(\ra)>0$ as $\lambda_{\shigh}'(0)<\ra<\lambda_{\shigh}'(1)$.
The same arguments show that $\ra<-\mathbb{E}_{\slow}\left[\ell\right]$
and $\lambda_{\slow}^{\star}(\ra)>0$. 
\end{proof}
\begin{proof}[Proof of Fact \ref{thm:ErrorProbSingleAgent-1}]
Consider the case $\Theta=\shigh$. As shown in Lemma \ref{lem:OptimalActions-1}
the probability that the agent makes a mistake is equal to the probability
that the LLR is below $\underline{L}(\act^{\shigh})$. Thus, Lemma \ref{lem:LLRDeviations}
allows us to characterize this probability explicitly: 
\[
\mathbb{P}_{\shigh}\left[a_{t}^{i}\neq\alpha^{\theta}\right]=\mathbb{P}_{\shigh}\left[R_{t}^{i}\leq\underline{L}(\act^{\shigh})\right]=\mathbb{P}_{\shigh}\left[R_{t}^{i}\leq o(t)\right]=\mathrm{e}^{-\lambda_{\shigh}^{\star}(0)\cdot t+o(t)}\,.
\]
An analogous argument yields that $\mathbb{P}_{\slow}\left[a_{t}^{i}\neq\alpha^{\theta}\right]=\mathrm{e}^{-\lambda_{\slow}^{\star}(0)\cdot t+o(t)}.$
By (\ref{eq:LegendreSym-1}) $\lambda_{\shigh}^{\star}(0)=\lambda_{\slow}^{\star}(0)\,.$ 
\end{proof}

\section{Many Agents\label{sec:AppendixBidirectional}}

Recall that we define for each $t$ the action $\actmin_{t}$ to be the lowest
action (i.e., having the lowest $\overline{L}(\alpha)$) that is taken
by any agent with positive probability at time $t$, and observe that
$\actmin_{t}$ is equal to $\alpha^{\slow}$ for all $t$ large enough.
We define
\[
G_{t}=\cap_{i=1}^{n}\cap_{\tau=1}^{t}\left\{ a_{\tau}^{i}=\actmin_{\tau}\right\} .
\]

\begin{proof}[Proof of Lemma \ref{lem:ThresholdRepresentation}]
Note first, that each agent chooses action $\actmin_{1}$ in the
first period if the likelihood ratio she infers from her first private
signal is at most $\overline{L}(\actmin_{1})$. Hence 
\[
G_{1}=\bigcap_{1\leq i\leq n}\{a_{1}^{i}=\actmin_{1}\}=\bigcap_{1\leq i\leq n}\{R_{1}^{i}\leq\overline{L}(\actmin_{1})\}.
\]
Thus the claim holds for $t=1$.
Assume now that all agents choose the action $\actmin_{\tau}$ up
to period $t-1$; that is, that $G_{t-1}$ has occurred, which is
a necessary condition for $G_{t}$. What would cause any one of them
to again choose $\actmin_{t}$ at period $t$? It is easy to see that
there will be some threshold $q_{t}^{i}$ such that, given $G_{t-1}$,
agent $i$ will choose $\actmin_{t}$ if and only if her private likelihood
ratio $R_{t}^{i}$ is lower than $q_{t}^{i}$. By the symmetry of
the equilibrium, $q_{t}^{i}$ is independent of $i$, and so we will
simply write it as $q_{t}$. It follows that 
\[
G_{t}=G_{t-1}\cap\bigcap_{1\leq i\leq n}\{R_{t}^{i}\leq q_{t}\}.
\]
Therefore, by induction, and if we denote $q_{1}=\overline{L}(\actmin_{\tau})$,
we have that 
\[
G_{t}=\bigcap_{\substack{\tau\leq t\\
1\leq i\leq n
}
}\{R_{\tau}^{i} \leq q_{\tau}\}.\qedhere
\]
\end{proof}
\begin{lem}
\label{prop:Recursive-Characterization-q}The threshold $q_{t}$ is
characterized by the recursive relation 
\begin{equation}
q_{t}=\overline{L}(\alpha_t^{\min})-(n-1)\cdot\log\frac{\mathbb{P}_{\shigh}\left[W_{t-1}^{1}\right]}{\mathbb{P}_{\slow}\left[W_{t-1}^{1}\right]}\quad\mbox{ and }\quad W_{t}^{i}=\bigcap_{1\leq\tau\leq t}\{R_{\tau}^{i}\leq q_{t}\}\,.\label{eq:q-t}
\end{equation}
\end{lem}
\begin{proof}
Agent $1$'s log-likelihood ratio conditional on $\cap_{i=1}^{n}W_{t-1}^{i}$
at time $t$ equals b
\begin{align*}
L_{t}^{1}=R_{t}^{1}+\log\frac{\mathbb{P}_{\shigh}\left[\cap_{i=2}^{n}W_{t-1}^{i}\right]}{\mathbb{P}_{\slow}\left[\cap_{i=2}^{n}W_{t-1}^{i}\right]}.
\end{align*}
Since the $W_{t-1}^{i}$'s are conditionally independent, we have
that 
\begin{align*}
L_{t}^{1}=R_{t}^{1}+\sum_{i=2}^{n}\log\frac{\mathbb{P}_{\shigh}\left[W_{t-1}^{i}\right]}{\mathbb{P}_{\slow}\left[W_{t-1}^{i}\right]}.
\end{align*}
Finally, by symmetry, all the numbers in the sum are equal, and 
\begin{align*}
L_{t}^{1}=R_{t}^{1}+(n-1)\cdot\log\frac{\mathbb{P}_{\shigh}\left[W_{t-1}^{1}\right]}{\mathbb{P}_{\slow}\left[W_{t-1}^{1}\right]}.
\end{align*}
Now, the last addend is just a number. Therefore, if we denote 
\begin{align}
q_{t}=\overline{L}(\alpha_t^{\min})-(n-1)\cdot\log\frac{\mathbb{P}_{\shigh}\left[W_{t-1}^{1}\right]}{\mathbb{P}_{\slow}\left[W_{t-1}^{1}\right]},\label{eq:t-k-1}
\end{align}
then 
\begin{align*}
L_{t}^{1}=R_{t}^{1}-q_{t}+\overline{L}(\alpha_t^{\min}),
\end{align*}
and $L_{t}^{1}\leq\overline{L}(\alpha_t^{\min})$ (and thus $a_{t}^{1}=\alpha_t^{\min}$)
whenever $R_{t}^{1}\leq q_{t}$. 
\end{proof}

\begin{lem}
\label{claim:q_t-pos}$q_{t}\geq\overline{L}(\alpha_{t}^{\min})$
for all $t$. 
\end{lem}
\begin{proof}
Let $F_{\shigh}$ and $F_{\slow}$ be the cumulative distribution functions
of a private log-likelihood ratio $\ell$, conditioned on $\Theta=\shigh$
and $\Theta=\slow$, respectively. Then it is easy to see that $F_{\shigh}$
stochastically dominates $F_{\slow}$, in the sense that $F_{\slow}(x)\geq F_{\shigh}(x)$
for all $x\in\mathbb{R}$.\footnote{To see this observe that for any non-decreasing function $h$, we have that $\mathbb{E}_{\shigh}[h(\ell)]=\mathbb{E}_{\slow}[h(\ell)\frac{\mathrm{d}\mu_{\shigh}}{\mathrm{d}\mu_{\slow}}(\ell)]=\mathbb{E}_{\slow}[h(\ell)\mathrm{e}^{\ell}]\geq \mathbb{E}_{\slow}[h(\ell)] \cdot\mathbb{E}_{\slow}[\mathrm{e}^{\ell}] = \mathbb{E}_{\slow}[h(\ell)] \cdot\mathbb{E}_{\slow}[\frac{\mathrm{d}\mu_{\shigh}}{\mathrm{d}\mu_{\slow}}(\ell)] = \mathbb{E}_{\slow}[h(\ell)],$ where the inequality follows from Chebyshev's sum inequality.} It follows that the joint distribution
of $(R_{\tau}^{i})_{\tau\leq t}$ conditioned on $\Theta=\shigh$ dominates
the same distribution conditioned on $\Theta=\slow$, and so $\mathbb{P}_{\shigh}\left[W_{t}^{1}\right]\leq\mathbb{P}_{\slow}\left[W_{t}^{1}\right]$.
Hence $q_{t}\geq\overline{L}(\alpha_{t}^{\min})$. 
\end{proof}
\begin{lem}
\label{claim:W_t-l}There is a constant $C>0$ such that $\mathbb{P}_{\slow}\left[W_{t}^{1}\right]\geq C$
for all $t$. 
\end{lem}
\begin{proof}
Since the events $W_{t}^{1}$ are decreasing, i.e.\  $W_t^1 \subseteq W_{t-1}^1$, we will prove the lemma
by showing that 
\[
\lim_{t\to\infty}\mathbb{P}_{\slow}\left[W_{t}^{1}\right]>0,
\]
which by definition is equivalent to 
\begin{align*}
\lim_{t\to\infty}\mathbb{P}_{\slow}\left[\cap_{\tau\leq t}\left\{ R_{\tau}^{i}\leq q_{\tau}\right\} \right]>0.
\end{align*}
Since $q_{t}\geq\overline{L}(\alpha_{t}^{\min})$, it suffices to
prove that 
\begin{align*}
\lim_{t\to\infty}\mathbb{P}_{\slow}\left[\cap_{\tau\leq t}\left\{ R_{\tau}^{i}\leq\overline{L}(\alpha_{\tau}^{\min})\right\} \right]>0.
\end{align*}

To prove the above, note that agents eventually learn $\Theta$, since
the private signals are informative. Therefore, conditioned on $\Theta=\slow$,
the limit of $R_{t}^{i}$ as $t$ tends to infinity must be $-\infty$.
Thus, with probability 1, for all $t$ large enough it does hold that
$R_{t}^{1}\leq\overline{L}(\alpha_{t}^{\min}).$ Since each of
the events $W_{t}^{1}$ has positive probability, and by the Markov
property of the random walk $R_{t}^{1}$, it follows that the event
$\cap_{\tau}\left\{ R_{\tau}^{i}\leq\overline{L}(\alpha_{\tau}^{\min})\right\} $
has positive probability. 
\end{proof}
\begin{lem}
\label{prop:The-limit-q-exists,}The limit $\beta=\lim_{t\to\infty}q_t/t$
exists, and 
\[
\beta=(n-1)\lambda_{\shigh}^{\star}(\beta)\,.
\]
\end{lem}
\begin{proof}
It follows immediately from Lemma~\ref{claim:W_t-l} and Lemma~\ref{prop:Recursive-Characterization-q} that 
\begin{equation}
\beta:=\lim_{t\to\infty}\frac{q_{t}}{t}=-(n-1)\lim_{t\to \infty}\frac{1}{t}\log\mathbb{P}_{\shigh}\left[W_{t-1}^{1}\right],\label{eq:q_t_i}
\end{equation}
provided that the limit exists. To show that this limit exists and to calculate it, let $\underline{\beta}=\lim\inf_{t\to\infty}q_{t}/t$. Since $W_{t}^{i}=\cap_{\tau=1}^{t}\left\{ R_{\tau}^{i}\leq q_{\tau}\right\} $,
it follows from Theorem~\ref{thm:large-deviation-2} that 
\[
\lim_{t\to\infty}\frac{1}{t}\log\mathbb{P}_{\shigh}\left[W_{t}^{i}\right]=-\lambda_{\shigh}^{\star}(\underline{\beta}),
\]
provided that $\underline{\beta}>\inf_{z}\lambda_{\shigh}'(z)$. But $\underline{\beta}\geq0$
(Lemma~\ref{claim:q_t-pos}), and so this indeed holds. The claim now follows from 
\eqref{eq:q_t_i}.
\end{proof}

\begin{lem}
\label{claim:q-bound}For any number of agents $n$ it holds that
$\rg<\mathbb{E}_{\shigh}\left[\ell\right]$. 
\end{lem}
\begin{proof}
Recall that $\lambda_{\shigh}^{\star}$ is strictly convex, and that $\lambda_{\shigh}^{\star}(\mathbb{E}_{\shigh}\left[\ell\right])=0$. Hence 
\begin{align*}
\lambda_{\shigh}^{\star}(\beta) & <\frac{\beta}{\mathbb{E}_{\shigh}\left[\ell\right]}\lambda_{\shigh}^{\star}(\mathbb{E}_{\shigh}\left[\ell\right])+\frac{\mathbb{E}_{\shigh}\left[\ell\right]-\beta}{\mathbb{E}_{\shigh}\left[\ell\right]}\lambda_{\shigh}^{\star}(0)\\
 & =\frac{\mathbb{E}_{\shigh}\left[\ell\right]-\beta}{\mathbb{E}_{\shigh}\left[\ell\right]}\lambda_{\shigh}^{\star}(0).
\end{align*}
Substituting $\beta/(n-1)$ for $\lambda_{\shigh}^{\star}(\beta)$  (which we can do by Lemma~\ref{prop:The-limit-q-exists,}) yields
\begin{align*}
    \frac{1}{n-1}\beta < \frac{\mathbb{E}_{\shigh}\left[\ell\right]-\beta}{\mathbb{E}_{\shigh}\left[\ell\right]}\lambda_{\shigh}^{\star}(0).
\end{align*}
Since $\lambda_{\shigh}^{\star}(0)<\mathbb{E}_{\shigh}\left[\ell\right]$ (Lemma \ref{claim:r^a-lambda}),
\begin{align*}
    \frac{1}{n-1}\beta <\mathbb{E}_{\shigh}\left[\ell\right]-\beta,
\end{align*}
or
\begin{align*}
    \frac{n}{n-1}\beta <\mathbb{E}_{\shigh}\left[\ell\right].
\end{align*}
Since, by \eqref{eq:rg}, $\rg = \frac{n}{n-1}\beta$, our proof is complete.
\end{proof}
We now turn to proving Theorem \ref{prop:groupthink}, which states
that conditioned on rational groupthink\textemdash that is, conditioned
on the event $G_{t}$\textemdash all agents have, with high probability,
a private LLR $R_{t}^{i}$ that strongly indicates the correct action.
In fact, we prove a stronger statement, which implies Theorem~\ref{prop:groupthink}: the private LLR is arbitrarily close to $\beta\cdot t$,
the asymptotic threshold for $R_{t}^{i}$ above which rational groupthink
ends. 
\begin{proof}[Proof of Theorem~\ref{prop:groupthink}]
We prove the theorem by showing a stronger statement. Namely, that for every $\epsilon>0$ it holds
that 
\[
\lim_{t\to\infty}\mathbb{P}_{\shigh}\left[R_{t}^{i}>t\cdot(\beta-\epsilon)\mbox{ for all }i\mid G_{t}\right]=1,
\]
where, as above, $\beta$ is the solution to $\beta=(n-1)\lambda_{\shigh}^{\star}(\beta)$.

By Theorem \ref{thm:large-dev-1} we know that 
\[
\lim{}_{t\to\infty}-\frac{1}{t}\log\mathbb{P}_{\shigh}\left[R_{t}^{i}\leq t\cdot(\beta-\epsilon)\right]=\lambda_{\shigh}^{\star}(\beta-\epsilon).
\]
Since $\lambda_{\shigh}^{\star}(\beta-\epsilon)>\lambda_{\shigh}^{\star}(\beta)$ it
follows that 
\[
\lim{}_{t\to\infty}-\frac{1}{t}\log\mathbb{P}_{\shigh}\left[A_{t}\right]=n\cdot\lambda_{\shigh}^{\star}(\beta-\epsilon)>n\cdot\lambda_{\shigh}^{\star}(\beta),
\]
where $A_{t}$ is the event $\left\{ R_{t}^{i}\leq t\cdot(\beta-\epsilon)\mbox{ for all }i\right\} $.
Since for $t$ high enough the event $A_{t}$ is included in $G_{t}$,
and since, by Lemma~\ref{prop:The-limit-q-exists,},
\[
\lim{}_{t\to\infty}-\frac{1}{t}\log\mathbb{P}_{\shigh}\left[G_{t}\right]=n\cdot\lambda_{\shigh}^{\star}(\beta),
\]
it follows that $\mathbb{P}_{\shigh}\left[A_{t}\mid G_{t}\right]$ decays
exponentially with $t$. Hence $\mathbb{P}_{\shigh}\left[A_{t}^{c}\mid G_{t}\right]\to_{t}1$,
which is the claim we set to prove. 
\end{proof}

\section{Early Period Mistake Probabilities}

We now prove Theorem \ref{thm:short-term}. We assume that each agent
$i$ observes a normal signal $s_{t}^{i}\sim\mathcal{N}(m_{\theta},n)$
with mean 
\[
m_{\Theta}=\begin{cases}
+1 & \text{ if }\Theta=\shigh\\
-1 & \text{ if }\Theta=\slow
\end{cases}
\]
and variance $n$. Note, that for any number of agents the precision of the joint signal
equals $1$, and thus the total information the group receives every
period is fixed, independent of $n$.

We assume that the prior belief assigns probability one-half to each
state $p_{0}=\nicefrac{1}{2}$ and that there are two actions $A=\{\slow,\shigh\}$
and each agent wants to match the state, as in the ``matching
the state'' example (\S\ref{subsec:Matching-the-State}). As
in the first period each agent bases her decision only on her own
private signal, she takes the action $\shigh$ whenever her signal $s_{1}^{i}$
is greater than $0$ and the action $\slow$ otherwise: 
\[
a_{1}^{i}=\begin{cases}
h & s_{1}^{i}>0\\
l & s_{1}^{i}\leq0
\end{cases}\,.
\]
The private likelihood of each agent after observing the first $t$
signals is given by 
\begin{align*}
R_{t}^{i} & =\log\frac{\prod_{\tau=1}^{t}\exp\left(-\frac{\left(s_{\tau}^{i}-1\right)^{2}}{2\,n}\right)}{\prod_{\tau=1}^{t}\exp\left(-\frac{\left(s_{\tau}^{i}+1\right)^{2}}{2\,n}\right)} =\frac{2}{n}\sum_{\tau=1}^{t}s_{\tau}^{i}\,.
\end{align*}
The probability that an agent takes the correct action $\Theta$ in
period 1 (conditional only on her own first period signal) is thus
given by 
\begin{align*}
\mathbb{P}_{\shigh}\left[\Theta=a_{1}^{i}\right] & =\mathbb{P}_{\shigh}\left[s_{1}^{i}\geq0\right] =1-\Phi\left(\frac{-m_{\shigh}}{\sqrt{n}}\right) =\Phi\left(\frac{1}{\sqrt{n}}\right)\,.
\end{align*}
By symmetry, $\mathbb{P}_{\slow}[a_{1}^{i}=\Theta]=\Phi(1/\sqrt{n})$
as well. Denote $\pi_{n}=\Phi\left(\frac{1}{\sqrt{n}}\right)$ and
by $N_{1}=|\{i\colon a_{1}^{i}=\shigh\}|$ the number of agents taking
the action $a_{1}^{i}=\shigh.$ Let $\kappa_{n}=\log(\pi_{n}/(1-\pi_{n}))$,
and note that $2/\sqrt{n}\geq\kappa_{n}\geq1/\sqrt{n}$.

As the action of each agent is independent, the LLR of agent $i$
at the beginning of period $2$ is given by 
\[
L_{2}^{i}=\frac{2}{n}(s_1^i+s_2^i)-(2\,N_{1}-n)\,\kappa_{n}-\text{sgn}(s_{1}^{i})\,\kappa_{n}\,.
\]
We define the public part of the LLR at the beginning of period $2$ as 
\[
L_{2}^{p}=(2\,N_{1}-n)\,\kappa_{n}\,.
\]
This is the LLR of an outside observer. We define the private part of the LLR as the remainder
\[
\hat{R}_{2}^{i}=L_2^i-L_2^p = \frac{2}{n}(s_{1}^{i}+s_2^i)-\text{sgn}(s_{1}^{i})\,\kappa_{n}\,.
\]
Let $\alpha_{m}$ be the action that the majority of the agents chose
in the first period (with $\alpha_{m}=\slow$ in case of a tie). Note
that $\alpha_{m}=\shigh$ iff $L_{2}^{p}>0$. Let $E_{t}$ be the event
that all agents take the first period majority action $\alpha_{m}$
in all subsequent periods up to time $t$, i.e., $a_{\tau}^{i}=\alpha_{m}$
for all $1<\tau\leq t$. 

\begin{proof}[Proof of Theorem~\ref{thm:short-term}]
We prove the theorem by showing that the probability of $E_{t}$ goes to one as the number of agents goes to infinity, i.e., 
\[
\lim_{n\to\infty}\mathbb{P}\left[E_{t}\right]=1\,.
\]
We in fact provide a quantitative statement and prove that $\mathbb{P}\left[E_{t}\right]\geq1-20\cdot t\cdot\sqrt{\frac{\log n}{n}}$ for all $n\geq 3$.

We first show that the the probability of the event $E_{2}$ that
all agents take the same action in period 2 goes to one. The LLR of
agent $i$ at the beginning of period $2$ is given by 
\begin{align*}
L_{2}^{i} & =\frac{2}{n}\sum_{\tau=1}^{2}s_{\tau}^{i}+(2\,N_{1}-n)\,\kappa_{n}-\text{sgn}(s_{1}^{i})\,\kappa_{n} = \hat{R}_{2}^{i}+L_{2}^{p}\,.
\end{align*}
To show that $E_{2}$ has high probability we show that with high
probability it holds that $L_{2}^{p}$, the public LLR induced
by the first period actions, is large (in absolute value) and that
the private beliefs are all small. Intuitively, this holds since both
are (approximately) zero mean normal, with $L_{2}^{p}$ having constant
variance and $\hat{R}_{2}^{i}$ having variance of order $1/\sqrt{n}$.
It will then follow that with high probability the signs of $L_{2}^{p}$
and $L_{2}^{i}$ are equal for all $i$, which is a rephrasing of
the definition of $E_{2}$.

Let $A_t$ be the event that all of the private signals in the first
$t$ periods have absolute values at most $M=4\sqrt{n\log n}$. Using
the union bound (over the agents and time periods), this happens except
with probability at most 
\[
\mathbb{P}\left[A_t^{c}\right]\leq t\cdot n\cdot\mathbb{P}\left[|s_{t}^{i}|>M\right]\leq t\cdot n\cdot2\cdot\Phi\left(-\frac{1}{2}M/\sqrt{n}\right);
\]
the 1/2 factor in the argument of $\Phi$ is taken to account for
the fact that the private signals do not have zero mean. Since $\Phi(-x)<e^{-\frac{x^{2}}{2}}$
for all $x<-1$, we have that 
\[
\mathbb{P}\left[A_t^{c}\right]\leq\frac{2\cdot t}{n}.
\]
Let 
\[
\hat{R}_{t}^{i}=\frac{2}{n}\sum_{\tau=1}^{t}s_{\tau}^{i}-\text{sgn}(s_{1}^{i})\,\kappa_{n}.
\]
Thus the event $A_t$ implies that for all $\tau \leq t$
\[
|\hat{R}_{\tau}^{i}|\leq\frac{2}{n}\cdot \tau\cdot M+\kappa_{n}\leq8\cdot \tau\cdot\sqrt{\frac{\log n}{n}}+\frac{2}{\sqrt{n}}\leq9\cdot \tau\cdot\sqrt{\frac{\log n}{n}}.
\]

Let $B_t$ be the event that the absolute value of the public LLR $L_{2}^{p}$
is at least $9\cdot t\cdot\sqrt{\frac{\log n}{n}}$; this is chosen
so that the intersection of $A_t$ and $B_t$ implies $E_{t}$. Conditioned
on $\Theta=\shigh$, the random variable $N_{1}$ has the unimodal binomial
distribution $\mathcal{B}(n,\pi_{n})$, which has mode $\lfloor(n+1)\cdot\pi_{n}\rfloor$.
The probability at this mode is easily shown to be at most $1/\sqrt{n}.$\emph{
}The same applies conditioned on $\Theta=\slow$. It follows that the
probability of $B_t^{c}$, which by definition is equal to the probability
that $|N_{1}-n/2|\leq\frac{1}{\kappa_{n}}9\cdot t\cdot\sqrt{\frac{\log n}{n}}$,
is at most $\frac{2}{\kappa_{n}}9\cdot t\cdot\sqrt{\frac{\log n}{n}}$
times the probability of the mode, or 
\[
\mathbb{P}\left[B^{c}\right]\leq\frac{2}{\kappa_{n}}9\cdot t\cdot\sqrt{\frac{\log n}{n}}\cdot\frac{1}{\sqrt{n}}\leq18\cdot t\cdot\sqrt{\frac{\log n}{n}}.
\]
Together with the bound on the probability of $A$, we have that 
\[
\mathbb{P}\left[A_t\text{ and }B_t\right]\geq1-20\cdot t\cdot\sqrt{\frac{\log n}{n}},
\]
and in particular 
\[
\mathbb{P}\left[E_{2}\right]\geq1-40\cdot\sqrt{\frac{\log n}{n}}.
\]

We now claim that $A_t\cap B_t$ implies $E_{t}$. To see this, note that
as $A_t\cap B_t$ implies $E_{2}$, the agents all observe at period 2
that no other agent has a strong enough signal to dissent with the
first period majority. This only strengthens their belief in the first
period majority, requiring them an even higher (in absolute value)
threshold than $L_{2}^{p}$ to choose another action; the formal proof
of this statement is identical to the proof of Lemma~\ref{claim:q_t-pos}.
But since, under the event $A_t\cap B_t$, each of their private LLRs
$\hat{R}_{\tau}^{i}$ is weaker than $L_{2}^{p}$ for all $\tau\leq t$,
they will not do so at period 3, or, by induction, in any of the periods
prior to period $t$. This completes the proof. 
\end{proof}

\section{Incomplete Observations Structures}

In this section we study the case that agent 1 observes agent 2's actions, but not vice versa. We prove Theorem~\ref{thm:unidirectional}, and moreover precisely calculate the error rate of agent 1, which allows us to compare it to the error rate in the bidirectional case. We think that this result is of independent interest.
\begin{thm}
\label{prop:Unidirectional}The probability that agent 1 makes
a mistake if she observes agent 2's actions unidirectionally satisfies
\[
e^\rightarrow_t = \mathbb{P}_{\theta}\left[a_{t}^{1}\neq\act^{\shigh}\right]=e^{-\ru\cdot t+o(t)}\,,
\]
where $\ru:=\ra+\min\left\{ \lambda_{\shigh}^{\star}(\ra),\lambda_{\slow}^{\star}(\ra)\right\} =\min\left\{ \lambda_{\slow}^{\star}(-\ra),\lambda_{\shigh}^{\star}(-\ra)\right\} $. 
\end{thm}

In the case of normal signals we can calculate $\ru$ exactly:
\begin{cor}
\label{cor:nor-uni}
Let $\mu_{\theta}$ be the normal distribution with mean $m_{\theta}$
and variance $\sigma^{2}>0$. In this case $\ru=\frac{25}{16}\ra$. 
\end{cor}

This implies that agent 1 learns as fast as she would learn if she
observed $9/16\approx56\%$ of agent 2's private signals, instead
of her actions.

\subsubsection*{Observing the last action}
To gain some intuition into unidirectional observations, let us first assume that agent 1 observes only
agent 2's last action $a_{t-1}^{2}$, rather than the entire history of $2$'s actions. That is, at time $t$ the information available to agent $1$ is $s_1^1,\ldots,s_t^1,a_{t-1}^2$, and the information available to agent $2$ is only $s_1^2,\ldots,s_t^2$.

Bayes rule yields that the LLR of agent 1 when agent 2 takes the action
$\act$ is given by 
\begin{equation}
L_{t}^{1}=R_{t}^{1}+I_{t}(a_{t-1}^{2})\,,\label{eq:LLRUnidirectional}
\end{equation}
where $I_{t}(a_{t-1}^{2})$ is the amount by which agent 1's log-likelihood
is shifted when she observes agent 2 take action $a_{t-1}^{2}$ in
period $t-1$: 
\[
I_{t}(\alpha):=\log\frac{\mathbb{P}_{\shigh}\left[a_{t-1}^{2}=\act\right]}{\mathbb{P}_{\slow}\left[a_{t-1}^{2}=\act\right]}\,.
\]
The next claim shows
that there are three different types of inference $I_{t}(\alpha)$
agent 1 can draw from agent 2's behavior. 
\begin{lem}
\label{prop:LLLRShift}The function $I_{t}(\alpha)$ satisfies 
\[
I_{t}(\alpha)=\begin{cases}
-\ra\cdot t+o(t) & \text{ if }\act=\act^{\slow}\\
+\ra\cdot t+o(t) & \text{ if }\act=\act^{\shigh}\\
o(1) & \text{ if }\act\notin\{\act^{\slow},\act^{\shigh}\}
\end{cases}\,.
\]
\end{lem}
This lemma follows simply from Fact~\ref{thm:ErrorProbSingleAgent-1}, which characterizes agent 2's autarky behavior:
When agent 2 takes a certainty action $\act\in\{\act^{\slow},\act^{\shigh}\}$
agent 1 believes that agent 2 has strong evidence for the state in
which agent 2's action is optimal. If agent 2 does not take a certainty
action $\act\notin\{\act^{\slow},\act^{\shigh}\}$ agent 1 believes that agent
2 must have gotten a sequence of very uninformative signals as she
knows that agent 2's belief is bounded away from certainty. As a consequence
the influence that agent 2's action has on agent 1's LLR $I_{t}(\act)$
vanishes for large $t$ in this case.

The fact, that the amount by which a full certainty action of agent
2 shifts agent 1's belief is asymptotically linear in the period $t$,
with slope equal to the rate $\ra$, follows as, by Fact~\ref{thm:ErrorProbSingleAgent-1}, the probability of a mistake in autarky vanishes at the rate $\ra$:
\begin{align*}
I_{t}(\act^{\slow}) & =\log\frac{\mathbb{P}_{\shigh}\left[a_{t-1}^{2}=\act^{\slow}\right]}{\mathbb{P}_{\slow}\left[a_{t-1}^{2}=\act^{\slow}\right]}=\log\mathbb{P}_{\shigh}\left[a_{t-1}^{2}=\act^{\slow}\right]-\log\mathbb{P}_{\slow}\left[a_{t-1}^{2}=\act^{\slow}\right]\\
 & =\log\left(e^{-\ra\cdot t+o(t)}\right)-o(1)\\
 & =-\ra\cdot t+o(t)\,.
\end{align*}

Intuitively, as agent 1 knows that agent 2, who acts in autarky, will
take a suboptimal action approximately with probability $\mathrm{e}^{-\ra\cdot t}$,
agent 1 shifts her LLR by approximately $-\ra\cdot t$ when she sees
that agent 2 chose $\alpha^{\slow}$, and shifts by $+\ra\cdot t$ when
she sees agent 2 chose $\alpha^{\shigh}$. When agent 1 sees agent 2 take
an action that is not optimal in either state she concludes that agent
2 is uninformed and ignores her action.

To calculate the probability of a mistake by agent $1$, let us first consider the case
of the good state. Recall that the LLR of agent 1 is the sum of the
LLRs of her private signals $R_{\tau}^{1}$ as well as the inference
$I_{t}(a_{t-1}^{2})$ she draws from agent 2's action 
\begin{align}
L_{t}^{1} & =R_{t}^{1}+I_{t}(a_{t-1}^{2})=\begin{cases}
R_{t}^{1}-\ra\cdot t+o(t) & \text{ if }a_{t-1}^{2}=\act^{\slow}\\
R_{t}^{1}+\ra\cdot t+o(t) & \text{ if }a_{t-1}^{2}=\act^{\shigh}\\
R_{t}^{1}+o(1) & \text{ if }a_{t-1}^{2}\notin\{\act^{\slow},\act^{\shigh}\},
\end{cases}\label{eq:InferenceUnidirectional}
\end{align}
where the second equality follows from Lemma~\ref{prop:LLLRShift}.
As shown in Lemma \ref{lem:OptimalActions-1}, agent 1 makes a mistake
in the good state (i.e., does not choose $\alpha^{\shigh}$) whenever her
likelihood is below $\underline{L}(a^{\shigh})$. Thus, when $a_{t-1}^{2}=\alpha^{\slow}$,
agent 1 does not choose $\alpha^{\shigh}$ whenever $R_{t}^{1}\leq\ra\cdot t+o(t)$.
We can estimate the probability of this event using Lemma \ref{lem:LLRDeviations}:
it is $\mathrm{e}^{-\lambda_{\shigh}^{\star}(\ra)\cdot t+o(t)}$. A similar
calculation for the other two cases yields 
\begin{align}
\mathbb{P}_{\shigh}\left[a_{t}^{1}\neq\act^{\shigh}\mid a_{t-1}^{2}=\act\right] & =\mathbb{P}_{\shigh}\left[L_{t}^{1}\leq\underline{L}(\act^{\shigh})\mid a_{t-1}^{2}=\act\right]\label{eq:unidir-cond-prob}\\
 & =\begin{cases}
\textrm{e}^{-\lambda_{\shigh}^{\star}(+\ra)\cdot t+o(t)} & \text{ if }\act=\act^{\slow}\\
\textrm{e}^{-\lambda_{\shigh}^{\star}(-\ra)\cdot t+o(t)} & \text{ if }\act=\act^{\shigh}\\
\textrm{e}^{-\lambda_{\shigh}^{\star}(0)\cdot t+o(t)} & \text{ if }\act\notin\{\act^{\slow},\act^{\shigh}\}
\end{cases}\,.\nonumber 
\end{align}
To calculate the overall probability of a mistake in state $\shigh$ we
calculate the probability with which the three above cases occur.

First, consider the case where agent
1 chooses a wrong action and agent 2 chooses the correct action $\alpha^{\shigh}$.
By Fact~\ref{thm:ErrorProbSingleAgent-1} the probability that agent
2 chooses the correct action $a_{t-1}^{2}=\act^{\shigh}$ satisfies 
\[
\mathbb{P}_{\shigh}\left[a_{t-1}^{2}=\act^{\shigh}\right]=1-\mathrm{e}^{-\ra+o(t)}
\]
As a consequence the probability that agent 1 chooses a wrong action
and agent 2 chooses the correct action equals 
\begin{align}
\label{eq:uni-case1}
\mathbb{P}_{\shigh}\left[a_{t}^{1}\neq\act^{\shigh}\text{ and }a_{t-1}^{2}=a^{\shigh}\right] 
&=\mathbb{P}_{\shigh}\left[a_{t}^{1}\neq\act^{\shigh}\mid a_{t-1}^{2}=a^{\shigh}\right]\times\mathbb{P}_{\shigh}\left[a_{t-1}^{2}=\act^{\shigh}\right]\nonumber\\
 & =\textrm{e}^{-\lambda_{\shigh}^{\star}(-\ra)\cdot t+o(t)}\left(1-\mathrm{e}^{-\ra\cdot t+o(t)}\right)\nonumber\\
 & =\textrm{e}^{-\lambda_{\shigh}^{\star}(-\ra)\cdot t+o(t)}\,.
\end{align}
The analysis of the other two cases (i.e., when agent 2 chooses $\alpha=\alpha^{\slow}$ or $\alpha \not\in \{\alpha^{\slow},\alpha^h\}$) are completed in the proof of the following lemma.
\begin{lem}
\label{prop:UnidirectionalLast}The probability that agent 1 makes
a mistake if she observes agent 2's last action unidirectionally satisfies
\[
\mathbb{P}\left[a_{t}^{1}\neq\act^{\Theta}\right]=e^{-\ru\cdot t+o(t)}\,,
\]
where $\ru:=\ra+\min\left\{ \lambda_{\shigh}^{\star}(\ra),\lambda_{\slow}^{\star}(\ra)\right\} =\min\left\{ \lambda_{\slow}^{\star}(-\ra),\lambda_{\shigh}^{\star}(-\ra)\right\} $. 
\end{lem}
\begin{proof}
Assuming that agent 1 only observes the last action of agent 2, we
would like to calculate $\mathbb{P}_{\shigh}\left[a_{1}^{t}\neq\act^{\shigh}\right]$.
We can write this as 
\begin{align}
\label{eq:uni1}
\lefteqn{\mathbb{P}_{\shigh}\left[a_{t}^{1}\neq\act^{\shigh}\right]}\nonumber\\
&=\mathbb{P}_{\shigh}\left[a_{t}^{1}\neq\act^{\shigh},a_{t-1}^{2}=\alpha^{\shigh}\right]+\mathbb{P}_{\shigh}\left[a_{t}^{1}\neq\act^{\shigh},a_{t-1}^{2}=\alpha^{\slow}\right]+\mathbb{P}_{\shigh}\left[a_{t}^{1}\neq\act^{\shigh},a_{t-1}^{2}\not\in\left\{ \alpha^{\shigh},\alpha^{\slow}\right\} \right].
\end{align}
We already calculated the first term in \eqref{eq:uni-case1}: it is equal to $\textrm{e}^{-\lambda_{\shigh}^{\star}(-\ra)\cdot t+o(t)}$.
To calculate the second term we write 
\begin{align*}
\mathbb{P}_{\shigh}\left[a_{t}^{1}\neq\act^{\shigh}\text{ and }a_{t-1}^{2}
=\alpha^{\slow}\right]
&=\mathbb{P}_{\shigh}\left[a_{t}^{1}\neq\act^{\shigh}\mid a_{t-1}^{2}=\alpha^{\slow}\right]\times\mathbb{P}_{\shigh}\left[a_{t-1}^{2}=\act^{\slow}\right]\\
&=\textrm{e}^{-\lambda_{\shigh}^{\star}(+\ra)\cdot t+o(t)}\times\mathbb{P}_{\shigh}\left[a_{t-1}^{2}=\act^{\slow}\right],
\end{align*}
where the second equality is an application of (\ref{eq:unidir-cond-prob}).
To estimate $\mathbb{P}_{\shigh}\left[a_{t-1}^{2}=\act^{\slow}\right]$ we
note that agent $2$ acts as in autarky, and therefore, by Lemma \ref{lem:LLRDeviations},
$\mathbb{P}_{\shigh}\left[a_{t-1}^{2}=\act^{\slow}\right]=\mathrm{e}^{-\lambda^{\star}(0)\cdot t+o(t)}=\mathrm{e}^{-\ra\cdot t+o(t)}$.
Hence 
\[
\mathbb{P}_{\shigh}\left[a_{t}^{1}\neq\act^{\shigh}\text{ and }a_{t-1}^{2}=\alpha^{\slow}\right]=\textrm{e}^{-(\lambda_{\shigh}^{\star}(\ra)+\ra)\cdot t+o(t)}.
\]
We are thus left with the estimation of the last addend, $\mathbb{P}_{\shigh}\left[a_{t}^{1}\neq\act^{\shigh}\text{ and }a_{t-1}^{2}\not\in\left\{ \alpha^{\shigh},\alpha^{\slow}\right\} \right]$.
To this end we note that 
\[
\mathbb{P}_{\shigh}\left[a_{t-1}^{2}\not\in\left\{ \alpha^{\shigh},\alpha^{\slow}\right\} \right]\leq\mathbb{P}_{\shigh}\left[R_{t}^{2}\leq\underline{L}(\alpha^{\shigh})\right]=\mathrm{e}^{-\ra\cdot t+o(t)},
\]
where the last equality is another consequence of Lemma \ref{lem:LLRDeviations}.
Therefore, by (\ref{eq:unidir-cond-prob}), 
\[
\mathbb{P}_{\shigh}\left[a_{t}^{1}\neq\act^{\shigh}\text{ and }a_{t-1}^{2}\not\in\left\{ \alpha^{\shigh},\alpha^{\slow}\right\} \right]=\mathrm{e}^{-2\ra t+o(t)}.
\]
We thus have that 
\[
\mathbb{P}_{\shigh}\left[a_{t}^{1}\neq\act^{\shigh}\right]=\textrm{e}^{-\lambda_{\shigh}^{\star}(-\ra)\cdot t+o(t)}+\textrm{e}^{-(\lambda_{\shigh}^{\star}(\ra)+\ra)\cdot t+o(t)}+\mathrm{e}^{-2\ra t+o(t)}.
\]
Recall that $\lambda_{\slow}^{\star}(\eta)=\lambda_{\shigh}^{\star}(-\eta)-\eta$
(by (\ref{eq:lambda-h-lambda-l})) and so $\lambda_{\shigh}^{\star}(\ra)+\ra=\lambda_{\slow}^{\star}(-\ra).$
Hence 
\[
\mathbb{P}_{\shigh}\left[a_{t}^{1}\neq\act^{\shigh}\right]=\textrm{e}^{-\lambda_{\shigh}^{\star}(-\ra)\cdot t+o(t)}+\textrm{e}^{-\lambda_{\slow}^{\star}(-\ra)\cdot t+o(t)}+\mathrm{e}^{-2\ra t+o(t)}.
\]
We show in Lemma~\ref{clm:2ra} below that $\lambda_{\shigh}^{\star}(-\ra)<2\ra$,
and likewise $\lambda_{\slow}^{\star}(-\ra)<2\ra$. Given this, the last
addend can be absorbed into the $o(t)$ term, and we have that 
\[
\mathbb{P}_{\shigh}\left[a_{t}^{1}\neq\act^{\shigh}\right]=\mathrm{e}^{-\ru\cdot t+o(t)},
\]
where 
\[
\ru=\min\left\{ \lambda_{\shigh}^{\star}(-\ra),\lambda_{\slow}^{*}(-\ra)\right\} =\ra+\min\left\{ \lambda_{\slow}^{\star}(\ra),\lambda_{\shigh}^{\star}(\ra)\right\} .
\]
By symmetry the same holds conditioned on $\Theta=\slow$, and so we have
shown that

\[
\mathbb{P}\left[a_{t}^{1}\neq\act^{\theta}\right]=\mathrm{e}^{-\ru\cdot t+o(t)}.
\]

This concludes the proof of Lemma~\ref{prop:UnidirectionalLast}.
\end{proof}

The next claim is used in the proof of the lemma above. Moreover, it shows that $\ru < 2\ra$: that is, for agent $1$ in this setting, learning from actions  is slower than learning from signals.
\begin{lem}
\label{clm:2ra}$\lambda_{\shigh}^{\star}(-\ra)<2\ra$ and $\lambda_{\slow}^{\star}(-\ra)<2\ra$. 
\end{lem}

\begin{proof}
We show the former; the proof of the latter is identical. To this
end, we first note that $-\ra>\lambda_{\shigh}'(1)$ (Claim \ref{claim:r^a-lambda}).
It thus follows that the maximum in 
\[
\lambda_{\shigh}^{\star}(-\ra)=\max_{z\geq0}\lambda_{\shigh}(z)+\ra z
\]
is also obtained in $(0,1)$, since the $z$ in which it is obtained
is the solution to $\lambda'(z)=-\ra$. Thus 
\[
\lambda_{\shigh}^{\star}(-\ra)=\max_{z\in(0,1)}\lambda_{\shigh}(z)+\ra z<\max_{z\in(0,1)}\lambda_{\shigh}(z)+\ra=2\ra.\qedhere
\]
\end{proof}

\subsubsection*{Observing all Actions Unidirectionally and the Proof of Theorem~\ref{prop:Unidirectional}}

We now return to the case that agent $1$ observes all of agent $2$'s past actions, while agent $2$ only observes her own signals. We show that in this case  the speed of learning is identical to the speed in the
case that she observes only the last action: 
\[
\mathbb{P}\left[a_{t}^{1}\neq\act^{\Theta}\right]=\mathrm{e}^{-\ru\cdot t+o(t)}.
\]
One direction is immediate: observing all actions can only reduce
the probability of error relative to observing the last action, and
so we know that 
\[
\mathbb{P}\left[a_{t}^{1}\neq\act^{\Theta}\right]\leq\mathrm{e}^{-\ru\cdot t+o(t)}.
\]
It thus remains to be shown that 
\[
\mathbb{P}\left[a_{t}^{1}\neq\act^{\theta}\right]\geq\mathrm{e}^{-\ru\cdot t+o(t)}.
\]
To show this we show that the probability of a smaller event already
satisfies this inequality. Specifically, we condition (without loss
of generality) on $\Theta=\shigh$ and would like to consider the case
that agent 2 chooses the wrong action $\alpha^{\slow}$ at all time periods
up to time t. As in Appendix~\ref{sec:AppendixBidirectional},  we define for each $t$ the action $\actmin_{\tau}$
to lowest (i.e., having the lowest $\overline{L}$) that is taken
by agent 2 with positive probability at time $t$. By the above, $\actmin_{\tau}$
is equal to $\alpha^{\slow}$ for all $t$ large enough. We then prove
the claim by showing that 
\begin{equation}
\mathbb{P}_{\shigh}\left[a_{t}^{1}\neq\act^{\shigh},\cap_{1\leq\tau\leq t}\{a_{\tau}^{2}=\actmin_{\tau}\}\right]=\mathrm{e}^{-\ru\cdot t+o(t)}.\label{eq:mistake-ballot}
\end{equation}
That is, we show that even when agent 1 observes agent 2 take the
wrong action at every period in which this is possible - even then
agent 1 gets it wrong with probability that is comparable to the probability
of mistake when observing only the last action. Denote by $E_{t}$
the event 
\[
E_{t}=\cap_{1\leq\tau\leq t}\left\{ a_{\tau}^{2}=\actmin_{\tau}\right\} .
\]
We first claim that 
\begin{equation}
\mathbb{P}_{\shigh}\left[E_{t}\right]=\mathrm{e}^{-\ra\cdot t+o(t)}\label{eq:E_t-h}
\end{equation}
and that 
\begin{equation}
\mathbb{P}_{\slow}\left[E_{t}\right]=e^{-o(t)},\label{eq:E_t-l}
\end{equation}
so that asymptotically this event has the same rate as the event $a_{t}^{2}=\alpha^{\slow}$,
for both possible values of $\Theta$. Given this, the analysis is
identical to the one carried out for the case of observing the last action only,
and likewise yields \eqref{eq:mistake-ballot}. It thus remains to
calculate the conditional rates of $E_{t}$, and in particular to
show that they are the same at the rates of the event $a_{t}^{2}=\alpha^{\slow}$.
The key insight from which this follows is the classical Ballot Theorem
\citep{bertrand1887solution}. It states that if $(X_{1},X_{2},\ldots)$
are i.i.d. random variables, and if $Y_{t}=\sum_{\tau=1}^{t}X_{\tau}$
then 
\[
\frac{1}{t}\mathbb{P}\left[Y_{t}\leq0\right]\leq\mathbb{P}\left[\cap_{\tau=1}^{t}\left\{ Y_{\tau}\leq0\right\} \right]\leq\mathbb{P}\left[Y_{t}\leq0\right],
\]
and so in particular the event that $Y_{t}\leq0$ has the same rate
as the event that $Y_{\tau}\leq0$ for all $\tau\leq t$. Instead
of using the Ballot Theorem, we use our Theorem~\ref{thm:large-deviation-2}.

Indeed, noting that the event $E_{t}$ can be written as 
\[
E_{t}=\cap_{1\leq\tau<t}\left\{ R_{\tau}^{2}\leq\overline{L}(\actmin_{\tau})\right\} .
\]
Thus, if we define $X_{t}=\ell_{t}$ and $y_{t}=\overline{L}(\actmin_{\tau})-L_{0}$
then $\lim_{t}y_{t}/t=0$ and Theorem \ref{thm:large-deviation-2}
yields the desired rates. This completes the proof of Theorem~\ref{prop:Unidirectional}.

\begin{proof}[Proof of Theorem~\ref{thm:unidirectional}]
Theorem~\ref{prop:Unidirectional} yields
\[
e^\rightarrow_t=e^{-\ru\cdot t+o(t)}\,.
\]
Thus, to complete the proof of the Theorem, we need to analyze $e^\leftrightarrow_t$, the probability of error in the bidirectional case. Indeed, it suffices to lower bound it, and show that its rate is lower than $\ru$.

We already lower bound $e^\leftrightarrow_t$ in our main results. There, we show that it is at least $e^{-\rg \cdot t +o(t)}$, where $\rg$, the rate of the groupthink event, conditioned on the good state, is given in \eqref{eq:rg} by $\rg=\frac{n}{n-1}\beta$, and where $\beta$ is the solution of the fixed point equation \eqref{eq:q}. In the case of $n=2$ agents, we get that
\[
\mathbb{P}_{\shigh}\left[a_{t}^{1} \neq \alpha^\Theta\right]\geq\mathrm{e}^{-2\beta\cdot t+o(t)}.
\]
and that $\beta$ is given by the fixed point equation $\beta=\lambda_{\shigh}^{\star}(\beta)$.

In the normal signal case this rate is about $1.37\ra$, and is in particular less than $\ru=\frac{25}{16}\ra\approx1.56\ra$. To complete the proof we need to show that for every signal distribution it still holds that $2\beta < \ru$.

Since $\ru=\min\left\{ \lambda_{\slow}^{\star}(-\ra),\lambda_{\shigh}^{\star}(-\ra)\right\} $,
in order to prove the claim we need to show that $2\cdot \beta<\lambda_{\shigh}^{\star}(-\ra)$;
the corresponding condition for the bad state will follow by the same
argument.

We consider two cases. If $\beta\geq\ra$ then $\lambda_{\shigh}^{\star}(\beta)\geq\lambda_{\shigh}^{\star}(0)$,
since $\beta=\lambda_{\shigh}^{\star}(\beta)$ and $\ra=\lambda_{\shigh}^{\star}(0)$.
By the monotonicity of $\lambda_{\shigh}^{\star}$ (Lemma~\ref{prop:lambda-star})
it then follows that $\beta\leq0$. But this is false since it
implies that $\beta=\lambda_{\shigh}^{\star}(\beta)\geq\lambda_{\shigh}^{\star}(0)>0$,
and so we have reached a contradiction.

Hence $\beta<\ra$, in which case $\lambda_{\shigh}^{\star}(-\beta)<\lambda_{\shigh}^{\star}(-\ra)$,
since $\lambda_{\shigh}^{\star}$ is strictly decreasing (Lemma~\ref{prop:lambda-star}).
Now, since $\beta=\lambda_{\shigh}^{\star}(\beta)$, we have that $2\cdot \beta=\beta+\lambda_{\shigh}^{\star}(\beta)=\lambda_{\slow}^{*}(-\beta)$,
where the last equality follows from the general fact (see Appendix
\ref{app:application-of-large-dev}) that $\lambda_{\slow}^{\star}(\eta)=\lambda_{\shigh}^{\star}(-\eta)-\eta$.
\end{proof}

\bigskip

\begin{proof}[Proof of Theorem~\ref{prop:observing-actions}]
Condition on $\Theta=\shigh$, with the other case admitting identical analysis. The probability that each of the agents $2,\ldots,n$ does not choose the correct action $\alpha^{\shigh}$ is $\gamma_t=e^{-\ra\cdot t+o(t)}$, by Fact~\ref{thm:ErrorProbSingleAgent-1}. Denote by $M_t$ the event that the majority of these agents did not take the action $\alpha^{\shigh}$. Since these actions are conditionally independent, by the Chernoff-Hoeffding bound 
\begin{align*}
    \mathbb{P}_{\shigh}[M_t] \leq \left(4\gamma_t(1-\gamma_t)\right)^{(n-1)/2} 
\end{align*}
and hence 
\begin{align*}
    \mathbb{P}_{\shigh}[M_t] \leq e^{-\frac{\ra t +o(t)-\log 4}{2}(n-1)}.
\end{align*}
Thus, if we take any $r < \ra/2$, we have that
\begin{align*}
    \mathbb{P}_{\shigh}[M_t] \leq e^{-(n-1)r \cdot t +o(t)}.
\end{align*}
The claim follows from the fact that agent 1's probability of mistake is lower than it would be if she were to (suboptimally) choose an action based on the majority of the actions of the others.
\end{proof}

\section{A Description of the Numerical Procedure}
\label{app:simulations}

A major challenge in computing behaviour numerically is that for $n$ agents the set of private histories at time $t$ equals $\mathbb{R} \times A^{t \times n}$, where the first component equals the private log-likelihood ratio and the second component equals the public history of actions taken by every agent until period $t$.
To simplify the exposition we focus on the case of two actions $|A|=2$.
In this case, for every public history of actions there exists a cut-off on the private LLR such that above the cut-off one action is taken and below the other action is taken.
A strategy that specifies behaviour at time $t$ thus corresponds to a vector of cut-offs with dimension $2^{(t-1) \times n}$. For example, for $10$ agents in period $10$ this space has dimension $2^{90} \approx 1.2 \times 10^{27}$. 
This ``curse of dimensionality'' makes the computation of equilibria challenging. Indeed, it has been shown in related settings that calculating the actions of Bayesian agents is  computationally hard \citep{pmlr-v99-hazla19a}.

We overcome this challenge by computing the agents' best response and beliefs for each realized path of actions using a nested Monte-Carlo simulation, rather than computing the whole equilibrium strategy. The main idea of our approach is to keep track of the agents' private LLRs, and the beliefs an outside observer would have about each agent's private LLR. As each agent's belief can be computed from these statistics, it suffices to keep track of these $n$ numbers and $n$ beliefs, independent of the time period. The caveat here is that the beliefs of an outside observer are distributions, which are infinite dimensional. To solve this issue we simply approximate these distributions by large, but finite samples.

%Given an sequence of actions taken by agent $i$, the rest of the agents make an inference regarding $i$'s private log-likelihood ratio, which then influences their own belief and action. This inference depends on the the agents' beliefs regarding $i$'s log-likelihood ratio: that is, the distribution of $i$'s private log-likelihood ratio, conditioned on the public history. We do not calculate this distribution exactly, but approximate it using a Monte-Carlo simulation. To each agent $i$ we associate a population of some fixed number $s$ of ``clones'' of $i$. To each clone we choose at random a sequence of private signals that is consistent with $i$'s public actions. We update this Thus each clone 

%More precisely, this Monte-Carlo simulation works as follows. 

%We first draw for each agent a random signal realization which we use to compute the action this agent takes in the initial period. We then compute for each agent the likelihood an outside observer assigns to the agent taking this action in each state by randomly drawing signals that are compatible with the agent's actions.
%These likelihoods can then be used to compute the agents beliefs along this sequence of actions.

\end{document}